\documentclass[english]{article}
\usepackage[T1]{fontenc}
\usepackage[latin9]{inputenc}
\usepackage{geometry}
\usepackage{amsmath}
\usepackage{amssymb}
\usepackage{dsfont}
\usepackage{mathtools}
\usepackage{xcolor}
\usepackage{cite}
\usepackage[hidelinks]{hyperref}

\DeclareMathOperator{\Tr}{Tr}
\DeclareMathOperator{\Sym}{Sym}

\makeatletter
\usepackage{braket}
\renewcommand\[{\begin{equation}}
\renewcommand\]{\end{equation}}

\let\mc=\mathcal

\makeatother

\usepackage{babel}
\begin{document}

\title{\bf{Target Space Entanglement Entropy}}

\author{Edward Mazenc \& Daniel Ranard}
\date{}

\maketitle

\begin{center}
    \small{\it{Department of Physics, Stanford University, \\
Stanford, CA 94305-4060, USA }}
\end{center}

\vspace{2cm}

\begin{abstract}
We define a notion of target space entanglement entropy. Rather than partitioning the base space on which the theory is defined, we consider partitions of the target space. This is the physical case of interest for first-quantized theories, such as worldsheet string theory. We associate to each subregion of the target space a suitably chosen sub-algebra of observables $\mathcal{A}$. The entanglement entropy is calculated as the entropy of the density matrix restricted to $\mc{A}$.  As an example, we illustrate our framework by computing spatial entanglement in first-quantized many-body quantum mechanics. The algebra $\mathcal{A}$ is chosen to reproduce the entanglement entropy obtained by embedding the state in the fixed particle sub-sector of the second-quantized Hilbert space.  We then generalize our construction to the quantum field-theoretical setting. 
\end{abstract}

\pagebreak

\tableofcontents

\pagebreak

\section{Introduction}

\begin{figure}[ht!]
    \includegraphics[width=\textwidth]{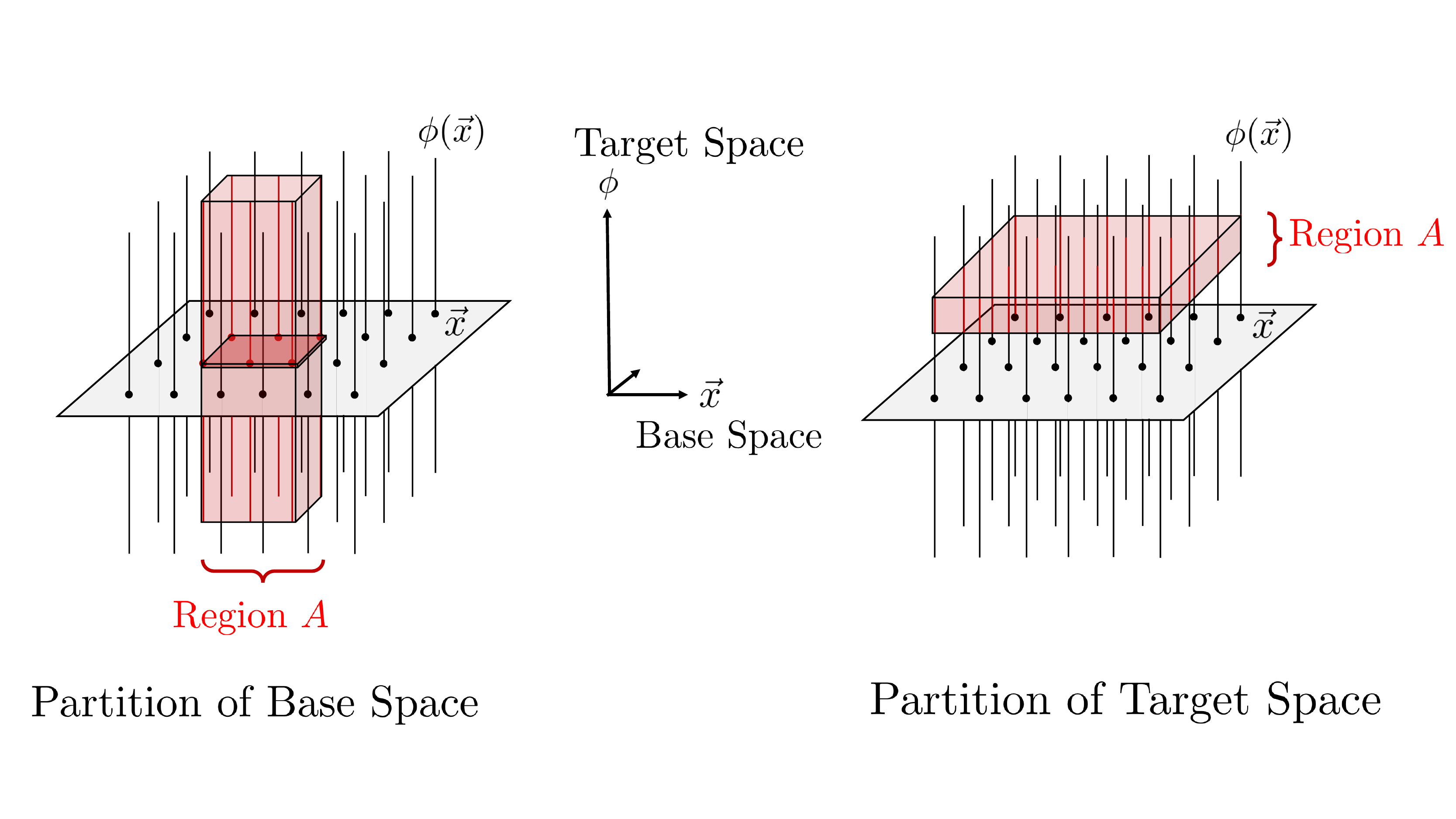}
    \centering
    \caption{\small{Distinction between partitions of the base space (left) versus target space (right). We illustrate the case of 2+1d scalar field theory for concreteness. At every point of the base space ($\cong \mathbb{R}^2$) labeled by $\vec{x}$, the local degree of freedom takes values in the target space ($\cong \mathbb{R}$) with coordinate $\phi$. Previous studies of entanglement entropy have focused on subregions of the base space where the values of $\vec{x}$ are restricted. Instead, we consider restrictions on the values of the field $\phi$.}}
    \label{fig:basevstarget}
\end{figure}

Previous studies of entanglement in field theory mostly address entanglement with respect to partitions of the base space \cite{casini2009entanglement,callan1994geometric}. Recall that in quantum field theory we speak of both the base space and the target space. For instance, in standard $d+1$ scalar field theory, the field $\phi(\vec{x})$ take values in the target space $\mathbb{R}$. The base space $\mathbb{R}^d$ parametrizes instead {\it{which}} degree of freedom we are speaking of: it labels the $\vec{x}\in \mathbb{R}^d$ of $\phi(\vec{x})$. Colloquially, we refer to the base space as the space the field ``lives on.'' To calculate the entanglement entropy of a spatial partition, we partition the base space $\mathbb{R}^d$; see Figure \ref{fig:basevstarget}.

Meanwhile, in promising theories of quantum gravity, a ``spatial'' partition may not be associated with a partition of the base space, but rather a partition of the target space.  For example, in first-quantized string theory, a spacetime subregion corresponds to a restriction of the embedding coordinates of the string. It is thus a partition of the target space, not a partition of the base worldsheet. 

Likewise, Matrix theory is a $0+1$ dimensional theory that describes, in a particular frame, quantum gravity in an 11-dimensional spacetime \cite{banks1997m}. In this case, the ``base space'' is nothing more than a point. Clearly, it is senseless to partition it. Instead, subregions of the physical spacetime correspond to a subspace of the moduli space of $D0$ branes \cite{maldacena2004exact,seiberg2006emergent}.

Motivated by these examples, the main goal of this paper is to define reduced density matrices and entanglement entropies of states with respect to subregions in target space.  The first challenge is that target space partitions do not correspond to tensor factorizations of the Hilbert space, whereas the usual framework for entanglement entropy hinges upon such a factorization.  We therefore leverage the powerful algebraic framework, which defines a reduced density matrix relative to a subalgebra of observables, and treats tensor factorizations as a special case.  This algebraic framework is by no means new, though it has only recently gained widespread traction in the high energy community via the work of \cite{casini2014remarks,harlow2017ryu,witten2018aps}, which offer excellent introductions to the subject. Our task therefore reduces to finding which subalgebra most accurately reflects an agent having access only to observables confined to the spatial subregion of interest. We then define target space entanglement entropy as the entropy of the state restricted to this subalgebra.

\begin{figure}[ht!]
    \includegraphics[scale=0.4]{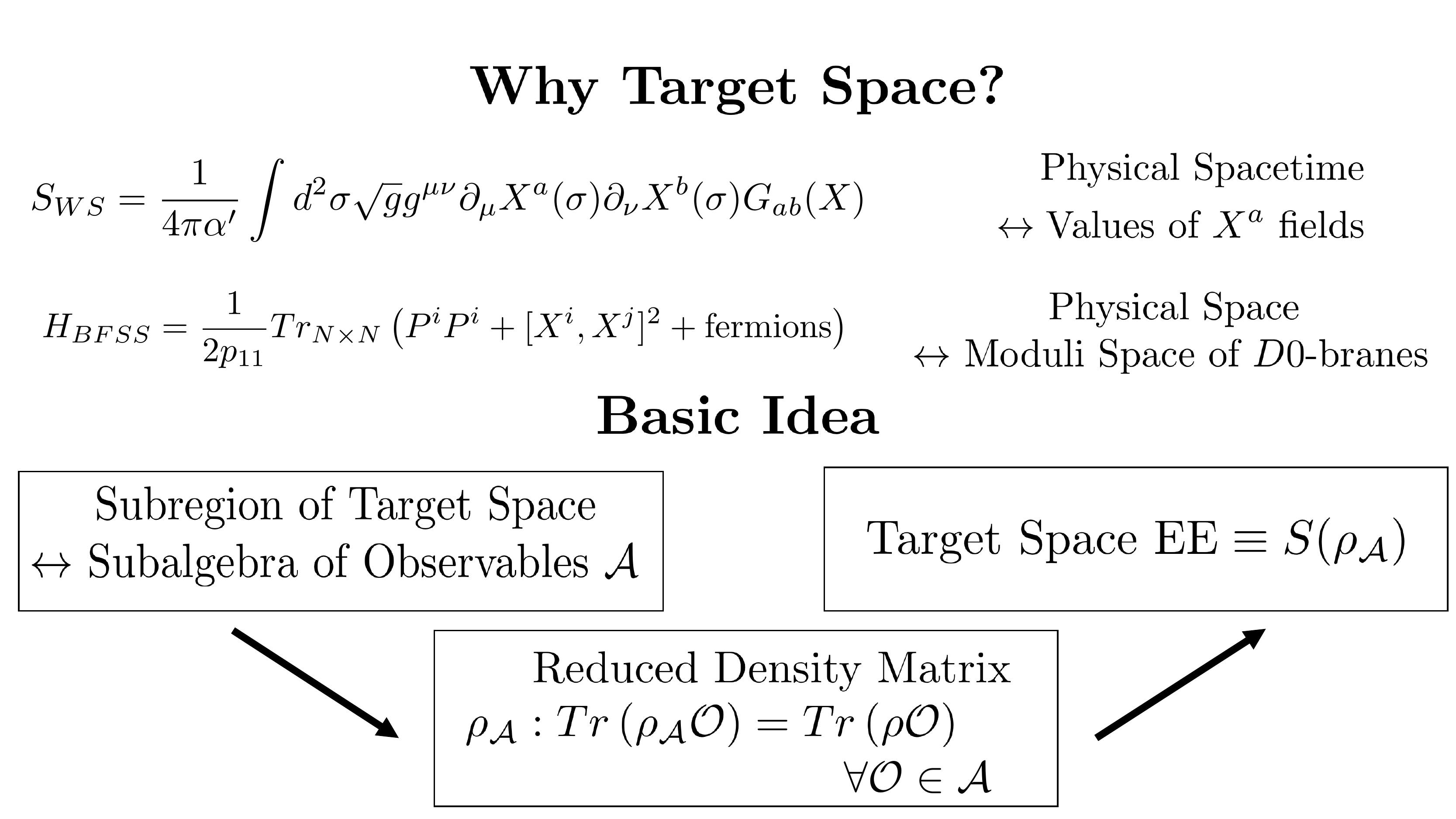}
    \centering
    \caption{\small{(Top) For many of our most promising theories of quantum gravity, such as worldsheeet string theory or the BFSS matrix quantum mechanics, the emergent physical spacetime of interest is encoded in the target space of the theory. (Bottom) To each subregion of the target space, we associate a particular subalgebra of observables $\mathcal{A}$. The reduced density matrix $\rho_{\mathcal {A}}$ is defined via restriction of the  state to the subalgebra $\mathcal{A}$. We define the target space entanglement entropy as the entropy of $\rho_{\mc{A}}$.}}
    \label{fig:mainidea}
\end{figure}

 A single particle on a line furnishes the simplest toy model. We may think of the position of the particle $x(t)$ as 0+1-dimensional QFT. The base space is a single point, and the target space $\mathbb{R}$ is the physical space the particle is moving in. To define a notion of spatial entanglement, we must partition the ``target space.'' First-quantized many body quantum mechanics provides the ideal testing ground for our proposed definition because we have a firm grasp on its second quantized formulation - plain old QFT - where we understand entanglement entropy well. Making a similar comparison in the string theory context would require the intricacies of string field theory (see \cite{balasubramanian2018remarks} for recent work directly in that context).  

In the non-relativistic case, our definition of entanglement entropy for the above quantum-mechanical system agrees with the the standard field theory definition. For relativistic quantum field theory, we find two ostensibly natural notions of locality and discuss their relative merits. An explicit computation for one-particle excited states shows to what extent the entanglement entropy associated to these different ``spatial'' partitionings can be compared. 

Further, we stress our framework is by no means limited to quantum mechanics. We generalize our construction to partitions of the target spaces of arbitrary sigma models and interacting field theories. We compute the entanglement entropy for a half (target) space partition in the the simplest example, a massive scalar field on two spatial lattice sites. 

We conclude by discussing the important role played by reparametrization invariance in theories such as worldsheet string theory and point out the limitations of our framework. We then apply the lessons learned in Section \ref{sec:axVSphix} to the case of 2d string theory and the ``baby cousin'' of the AdS/CFT correspondence, the holographic $c=1$ matrix model. There, we face multiple notions of emergent locality \cite{seiberg2006emergent}, and we sketch the possible role of different factorizations of the Hilbert space. 
 
\section{Preview: Particle on a line \& the need for subalgebras} \label{sec:preview}

To handle partitions of target space, we will need the algebraic framework for entanglement entropy.  However, let us motivate it further by exploring in more detail the example of particle on a line.

The Hilbert space $L^2(\mathbb{R}) = \textrm{span}\{|x\rangle : x \in \mathbb{R}\}$ may also be considered as the Hilbert space of a 0+1-dimensional QFT, where the base space is a single point, and $L^2(\mathbb{R}) = \textrm{span}\{|\phi\rangle : \phi \in \mathbb{R}\}$ is the space of field values $\phi$ at that point.  From the latter perspective, we will call $\mathbb{R}$ the ``target space.''  Alternatively, from the perspective of quantum mechanics on the line, $\mathbb{R}$ is simply the space on which the particle lives.  

We can partition the target space $\mathbb{R}$ into a region $A$ and its complement $\bar{A}$; for instance, we might choose half-spaces $A \equiv \{ x : x \leq x_0\}$ and $\bar{A} \equiv \{ x : x > x_0\}$.  This bi-partition induces a decomposition of the Hilbert space into a direct sum, 
\begin{equation} \label{eqn:particle-line-direct-sum}
    \mathcal{H}=L^2(A \cup \bar{A})=\mathcal{V}_{A}\oplus\mathcal{V}_{\bar{A}}
\end{equation}
where 
\begin{align}
\mathcal{V}_{A} & \equiv \textrm{span}\{\ket{x} : x\in A\}, \\
 \mathcal{V}_{\bar{A}}  & \equiv \textrm{span}\{\ket{x} : x\in \bar{A}\}. \nonumber
\end{align}

We emphasize that this decomposition is \textit{not} a tensor factorization.  Therefore one might wonder how to define a subsystem, a partial trace and reduced density matrix, or an entanglement entropy.  To proceed, we thus review a more general notion of subsystems, based on sub-algebras rather than tensor factors.

\section{Review of the algebraic definition of entanglement entropy}

Traditionally, one defines the entanglement of a state $|\psi\rangle \in \mathcal{H}$ relative to some bi-partition of the Hilbert space $\mathcal{H} = \mathcal{H}_{A} \otimes \mathcal{H}_{\bar{A}}$, where we have divided the degrees of freedom into subsystem $A$ and its complement $\bar{A}$.  We will consider pure states on $\mathcal{H}$, in which case the entanglement can be quantified by the the von Neumann entanglement entropy of the reduced state $\rho_{A} = \Tr_{\bar{A}}(|\psi\rangle \langle \psi | )$.

There are many ways to factorize a Hilbert space $\mathcal{H}$ as $\mathcal{H} = \mathcal{H}_{A} \otimes \mathcal{H}_{\bar{A}}$, and different factorizations may be appropriate for different purposes.  Given a factorization, it is natural to consider the algebra $\mathcal{A}$ of operators local to $A$, i.e.\ operators of the the form $O_A \otimes \mathds{1}_{\bar{A}}$.   These operators represent the observables and operations available to an observer confined to subsystem $A$.

Even without a factorization of the Hilbert space, we can still choose a sub-algebra of ``accessible'' observables $\mathcal{A}$ and use this to \textit{define} the subsystem $A$.   Recall that an algebra $\mathcal{A}$ of operators on a Hilbert space $\mathcal{H}$ is a subset $\mathcal{A} \subset L(\mathcal{H})$ where $L(\mathcal{H})$ denotes the space of all linear operators; here we consider finite-dimensional ``von Neumann algebras,'' required to be closed under addition, multiplication, scaling, and Hermitian conjugation.\footnote{Though we often refer to it as an ``algebra of observables,'' not all of the elements need be Hermitian. Here we will require an algebra to include the identity element. There are additional subtleties in infinite dimensions that we will not immediately discuss, though the results presented above all hold for finite direct sums of Type $I$ factors.  See Section \ref{sec:finiteness} for more discussion of infinite dimensions.}   By identifying \textit{any} subalgebra $\mathcal{A} \subset L(\mathcal{H})$ with an abstract ``subsystem,'' we generalize the notion of a subsystem beyond tensor factors.

We review a few key facts about algebras of operators. The most important theorem is that given any algebra $\mathcal{A} \subset L(\mathcal{H})$, there exists a decomposition of the Hilbert space as a direct sum of tensor products,
\begin{equation} \label{eqn:decomposition}
\mathcal{H} = \bigoplus_i \mathcal{H}_{A,i} \otimes \mathcal{H}_{\bar{A},i} 
\end{equation}
such that the operators $\mc{O}_A \in \mathcal{A}$ are precisely those which take the form
\begin{equation}
\mc{O}_A = \sum_i \mc{O}_{A,i} \otimes \mathds{1}_{\mathcal{H}_{\bar{A},i}}
\end{equation}
for some $\mc{O}_{A,i} \in L(\mathcal{H}_{A,i})$.  This follows from a pedestrian version of the Artin-Wedderburn theorem.  Schematically, we can write
\begin{equation}
\mathcal{A} = \bigoplus_i  L(\mathcal{H}_{A,i}) \otimes \mathds{1}_{\mathcal{H}_{\bar{A},i}}.
\end{equation}

In the case that the above sum has only one term, $\mathcal{A}$ is called a ``factor,'' and indeed the Hilbert space tensor factorizes as $\mathcal{H}_A \otimes \mathcal{H}_{\bar{A}}$.  In infinite dimensions, the existence of this tensor factorization hinges upon the ``Type'' of algebra; see Section \ref{sec:finiteness} for more.  

Given an algebra $\mathcal{A} \subset L(\mathcal{H})$, an important related algebra is the \textit{commutant} $\mathcal{A}' \subset L(\mathcal{H})$, the set of operators that commute with all the operators on $\mathcal{A}$.  Given the above decomposition (\ref{eqn:decomposition}), Schur's lemma allows us to easily write down the commutant, which takes the form
\begin{align}
\mathcal{A}' = \bigoplus_i   \mathds{1}_{\mathcal{H}_{A,i}} \otimes L(\mathcal{H}_{\bar{A},i}).
\end{align}
We can also define the center $Z(\mathcal{A}) \equiv \mathcal{A} \cap \mathcal{A}'$, the set of operators on $\mathcal{A}$ that commute with all operators on $\mathcal{A}$.  The center may be expressed as
\begin{align} \label{eqn:minimal-projectors}
Z(\mathcal{A}) = \textrm{span}\{\Pi_i\}_i 
\end{align}
where $\Pi_i$ are the projectors onto the direct sum sectors $\mathcal{H}_{A,i} \otimes \mathcal{H}_{\bar{A},i}$. In practice, we often start out with $\mc{A}$, then determine the minimal projectors spanning its center. This in turn allows us to actually find the block decomposition of the Hilbert space laid out in Eqn.\ \ref{eqn:decomposition}. Note that when there is only one sector, i.e.\ $\mathcal{A}$ is a factor, the center contains only multiples of the identity.  

With these ingredients in hand, we can easily define the reduced density matrix with respect to a sub-algebra.  Say we have a state $\rho$ and want to define a reduced density state $\rho_A$ with respect to $\mathcal{A}$. First recall that for an ordinary tensor factorization $\mathcal{H} = \mathcal{H}_A \otimes \mathcal{H}_{\bar{A}}$, the reduced state $\rho_A$ can be defined as the unique state on $\mathcal{H}_A$ such that $\Tr(\rho_A \mc{O}_A) = \Tr(\rho \mc{O}_A)$ for all $\mc{O}_A \in L(\mathcal{H})$.  With that definition, one can show $\rho_A$ is given by the familiar partial trace.  

Analogously, for the case of an algebra, we will define $\rho_{\mc{A}}$ to be the unique element of $\mathcal{A}$ such that 
\begin{equation}
\Tr(\rho_{\mc{A}} \mc{O}_A) = \Tr(\rho \mc{O}_A)
\end{equation} 
for all $\mc{O}_A \in \mathcal{A}$.  Given the decomposition of Eqn.\ \ref{eqn:decomposition}, it turns out one can easily express $\rho_{\mc{A}}$ by using partial traces on each sector.  Let  
\begin{align} \label{eqn:probabilities}
p_i &  \equiv \Tr (\Pi_i \rho \Pi_i ), \\
\rho_i & \equiv \frac{1}{p_i} \Pi_i \rho \Pi_i.
\end{align}
Then one can show $\rho_A$ must be given by
\begin{equation} \label{eqn:RDM}
\rho_{\mc{A}} = \sum_i  p_i \Tr_{\bar{A},i}(\rho_i) \otimes \frac{\mathds{1}_{\mathcal{H}_{\bar{A},i}}}{\textrm{dim}(\mathcal{H}_{\bar{A},i})},
\end{equation}
where $\rho_i$ is a state living on the $i$'th sector $\mathcal{H}_{A,i} \otimes \mathcal{H}_{\bar{A},i}$.  The partial traces on each sector are well-defined because each sector factorizes individually.

To define the entanglement entropy of $\rho$, we further consider the state
\begin{equation}
\tilde{\rho}_{\mc{A}} = \sum_i  p_{i} \Tr_{\bar{A},i}(\rho_i)
\end{equation}
on the Hilbert space $\mathcal{H}_A \equiv \bigoplus_i \mathcal{H}_{A,i}$, where we have simply stripped off the identity factors.  Then we define the entanglement entropy of $\rho$ with respect to $\mathcal{A}$ as the ordinary von Neumann entropy of the state $\tilde{\rho}_A$ on the Hilbert space $\mathcal{H}_A$.\footnote{This definition differs from the naive definition $ \Tr_{\mc{H}}\left(\rho_{\mc{A}} \log \rho_{\mc{A}} \right)$ by the term $\Delta S = \sum_{i}p_{i}log \left({\textrm{dim}(\mathcal{H}_{\bar{A},i})}\right) $. To reproduce the standard entropy for the case of a factor, we must use the definition outlined in the main text. See Appendix A.7.2 of \cite{harlow2017ryu} or more broadly \cite{ohya2004quantum}. To avoid any confusion: throughout the text, when we refer to the von Neumann entropy of $\rho_{\mc{A}}$, we really mean $S(\rho,\mc{A})$, or equivalently $S(\tilde{\rho}_{\mc{A}})$.}  That is, we define
\begin{align} \label{eqn:algebraic-EE}
S(\rho,\mathcal{A}) & \equiv S(\tilde{\rho}_{\mc{A}})  \\
& = S\left(\sum_i  p_i \Tr_{\bar{A},i}(\rho_i) \right) \\
& = - \sum_i p_i \log(p_i) + \sum_i p_i S(\rho_i)\\
& \equiv S(\rho,\mathcal{A})_{classical} + S(\rho,\mathcal{A})_{quantum}.
\end{align}
We find that the entanglement entropy breaks into two pieces, a ``classical'' piece and a ``quantum'' piece. $S(\rho,\mc{A})_{classical}$ is the Shannon entropy of the classical probability distribution $\{p_i\}$ over the different blocks (often referred to as ``superselection sector''). $S(\rho,\mathcal{A})_{quantum}$ on the other hand is the weighted sum of the quantum von Neumann entropy of the reduced density matrices $\rho_i$ within each block \cite{casini2014remarks,lin2018comments}.

Another simple way to define $S(\rho,\mathcal{A})$ is to embed $\mathcal{H}$ into an extended Hilbert space $\mathcal{H}_{ext}$ which \textit{does} have a tensor factorization.  In particular, we define
\begin{align}
\mathcal{H}_A & \equiv \bigoplus_i \mathcal{H}_{A,i},  \\
\mathcal{H}_{\bar{A}} & \equiv \bigoplus_i \mathcal{H}_{\bar{A},i} , \nonumber
\end{align}
so that we can define the extended Hilbert space
\begin{align} \label{eqn:extendedHS}
\mathcal{H}_{ext} & \equiv \mathcal{H}_A \otimes \mathcal{H}_{\bar{A}} \nonumber \\
& =  \bigoplus_{i,j} \mathcal{H}_{A,i} \otimes  \mathcal{H}_{\bar{A},j} \nonumber \\
& \supset  \bigoplus_i \mathcal{H}_{A,i} \otimes \mathcal{H}_{\bar{A},i} = \mathcal{H}.
\end{align}
Therefore we can also view the state $\rho$ on $\mathcal{H}$ as a state $\rho_{ext}$ on the extended Hilbert space $\mathcal{H}_{ext}$. Then $S(\rho,\mathcal{A})$ is then precisely the ``ordinary'' entanglement entropy obtained by taking the partial trace of $\rho$ over $\mc{H}_{\bar{A}}$ and then computing the von Neumann entropy of this reduced density matrix\footnote{Readers familiar with the literature on entanglement entropy in gauge theory might object that the ``extended Hilbert space'' and algebraic definitions famously disagree. However, the extended Hilbert space construction in gauge theory differs from the one in Eq. \ref{eqn:extendedHS}. 


}, i.e.\
\begin{align} \label{eqn:algebraic-EE-ext}
S(\rho,\mathcal{A}) =   \Tr_{\mc{H}_{A}}\left(\rho_A \log \rho_A \right) \\
 \text{with} \ \rho_A \equiv  \Tr_{\mc{H}_{\bar{A}}} \left( \rho_{ext} \right) \nonumber
\end{align}

In the case that $\mathcal{A}$ corresponds to the set of operators on a tensor factor, the quantity $S(\rho,\mathcal{A})$ agrees with the standard von Neumann entanglement entropy.  However, beyond agreement with the ``ordinary'' case, what motivates this definition of  $S(\rho,\mathcal{A})$?  

We might first ask the motivation for ordinary von Neumann entanglement entropy.  Besides proving a useful tool for analyzing field theories and many-body physics, the entanglement entropy affords several operational or information-theoretic interpretations.  For instance, the von Neumann entanglement entropy between subsystems $A$ and $\bar{A}$ also equals the ``distillable entanglement,'' the number of Bell pairs that can be distilled by observers on $A$ and $\bar{A}$ using only local operations on $A, \bar{A}$ and classical communication.  The entanglement entropy with respect to an algebra affords an analogous interpretation as distillable entanglement, where observers on $A$ and $\bar{A}$ are restricted to using operations associated to $\mathcal{A}$ and $\mathcal{A}'$, respectively.  However, it turns out the distillable entanglement is equal to the quantum piece $S(\rho,\mathcal{A})_{quantum}$ alone \cite{schuch2004nonlocal,Verstraete2016,soni2016aspects}.  See Equation \ref{eqn:apparatus-interaction} for an elaboration of the operational interpretation.

\section{Algebraic entanglement entropy for first-quantized systems} \label{sec:algebraic-EE}

We first apply the algebraic definition of entanglement entropy to our example of a particle on a line, spelling out the details on this first pass. We then proceed more generally to the first-quantized quantum mechanics of $N$ particles.  Later in Section \ref{sec:Fock}, we will confirm that our framework gives the same spatial entanglement entropy had we instead embedded the first-quantized, $N$-particle Hilbert space into the $N$-particle sector of a second-quantized Fock space, and defined the entanglement entropy using the tensor factorization associated to the Fock space. Nonetheless, we develop the algebraic approach as a general tool, applicable even when no obvious second-quantized theory exists.

\subsection{Single-particle warm-up}

We return to the particle on a line, introduced in Section \ref{sec:preview}.  The Hilbert space is simply $\mathcal{H} = L^2(\mathbb{R})$, and we can think of $\mathbb{R}$ alternately as the space on which the particle lives, or the target space of a 0+1-dimensional QFT.  For this section, we primarily adopt the language of the former. 

Partitioning the line $\mathbb{R}$ into a region $A \subset \mathbb{R}$ and its complement $\bar{A}$, we obtain the decomposition of Eqn.\  \ref{eqn:particle-line-direct-sum},
\begin{equation} 
    \mathcal{H}=\mathcal{V}_{A}\oplus\mathcal{V}_{\bar{A}},
\end{equation}
where $\mathcal{V}_{A} = \textrm{span}\{\ket{x} : x\in A\}$ and likewise $\mathcal{V}_{\bar{A}}  = \textrm{span}\{\ket{x} : x\in \bar{A}\}$.  

Now we choose an algebra $\mathcal{A} \subset L(\mathcal{H})$ to associate to the region $A$.  We propose the following algebra,
\begin{equation}
   \mathcal{A}= \bigg \langle \{ \ket{x}\bra{x'}: x,x'\in A \} \cup \mathds{1}_{\mathcal{H}} \bigg \rangle.
\end{equation}

The angular brackets denote ``the algebra generated by,'' i.e.\ the algebra of all operators generated by addition, multiplication, and scaling of the operators within the brackets.  To physically motivate this choice, note the Hermitian operators in $\mathcal{A}$ correspond to what observers situated in the region $A$ of the line could measure.  Including the identity is crucial. Physically, it corresponds to the fact that an observer should be able to act trivially on the system.  

It will also be useful to define the projector 
\begin{equation}
    \Pi_{A}=\int_{x\in A} dx\ket{x}\bra{x},
\end{equation}
which acts on the subspace $\mathcal{V}_A$ as the identity $\mathds{1}_{A}$.  We denote the orthogonal complement as $ \Pi_{\bar{A}}=\mathds{1}_{\bar{A}}$.

Written in the position basis, with the basis partitioned into elements in $A$ and $\bar{A}$, all operators $\mathcal{O} \in \mathcal{A}$ take the following form 
\begin{equation} \label{eqn:single-particle-algebra-block}
    \mathcal{O}=\left(\begin{array}{cc}
\mathcal{O}_{A} & 0\\
0 & c_{0}\mathds{1}_{\bar{A}}
\end{array}\right)
\end{equation}
where $c_{0}$ is an arbitrary constant and 

\begin{eqnarray*}
\mathcal{O}_{A} & = & \Pi_{A}\mathcal{O}\Pi_{A}\\
 & = & \int_{x,x'\in A} dx\,dx'\,\mathcal{O}(x,x')\ket{x}\bra{x'}.
\end{eqnarray*}

To analyze the structure of this algebra, we compute the commutant $\mathcal{A}'$, again the set of operators that commute with all those in $\mathcal{A}$. By Schur's Lemma, $\mathcal{A}'$ is given by all operators of the form 
\begin{equation} \label{eqn:single-particle-algebra-commutant-block}
 \mathcal{O}'=\left(\begin{array}{cc}
c_{1}\mathds{1}_{A} & 0\\
0 & \mathcal{O}_{\bar{A}}
\end{array}\right)
\end{equation}
with $c_{1}$ some other arbitrary constant.  Thus we could also denote $\mathcal{A}'$ as the algebra $\bar{\mathcal{A}}$ corresponding to the complementary region $\bar{A}$, with analogous definition
\begin{equation}
    \mathcal{A}'=\bar{\mathcal{A}} \equiv \bigg \langle \{\ket{x}\bra{x'}: x,x' \in \bar{A} \} \cup \mathds{1}_{\mc{H}} \bigg \rangle.
\end{equation}
Hence the center $\mathcal{Z}=\mathcal{A}\cap\mathcal{A}'$
is given by 
\begin{equation}
    \mathcal{Z}= \bigg \langle \int_{x\in A} dx \ket{x}\bra{x}\cup \mathds{1}_{\mc{H}} \bigg \rangle = \textrm{span}\{\Pi_A, \Pi_{\bar{A}}\}.
\end{equation}
The center being non-trivial simply reflects the fact that $\mathcal{A}$ does not induce a simple tensor factorization.

As guaranteed by the theorem of Eqn.\  \ref{eqn:decomposition}, the algebra $\mathcal{A}$ induces a decomposition of the Hilbert space.  The decomposition is apparent from the form of $\mathcal{A}$ and $\mathcal{A}'$ in Eqns. \ref{eqn:single-particle-algebra-block}, \ref{eqn:single-particle-algebra-commutant-block} above.  We have
\begin{align} \label{eqn:single-particle-decomposition}
\mathcal{H} & = \bigoplus_{i=0,1} \mathcal{H}_{A,i} \otimes \mathcal{H}_{\bar{A},i}.
\end{align}
where
\begin{align}
\mathcal{H}_{A,0}  & = \mathbb{C} \\
\mathcal{H}_{\bar{A},0}  & = \mathcal{V}_{\bar{A}} = \textrm{span}\{|x\rangle  : x \in \bar{A}\}  \\
\mathcal{H}_{A,1}  &  = \mathcal{V}_{A} = \textrm{span}\{|x\rangle  : x \in A\}  \\
\mathcal{H}_{\bar{A},1}  & = \mathbb{C}
\end{align}
so that
\begin{align}
\mathcal{H} & =  (\mathbb{C} \otimes \mathcal{V}_{\bar{A}}) \oplus (\mathcal{V}_A \otimes \mathbb{C})\\
& = \mathcal{V}_A \oplus \mathcal{V}_{\bar{A}},
\end{align}
recovering Eqn.\  \ref{eqn:particle-line-direct-sum}.

The decomposition here is slightly trivial, because the Hilbert spaces $\mathcal{H}_{A,0}$ and $\mathcal{H}_{\bar{A},1}$ happen to be the trivial space $\mathbb{C}$. Each sector corresponds to the number of particles in $A$.  For example, $\mathcal{H}_{A,1} \otimes \mathcal{H}_{\bar{A},1}$ is the sector where the particle is within $A$. It is the tensor product of $\mathcal{H}_{A,1}$, the space of wavefunctions on $A$, with the trivial space $\mathcal{H}_{\bar{A},1}$, whose single ray represents the state of $\bar{A}$ with zero particles.  Likewise, we can think of the sector $\mathcal{H}_{A,0} \otimes \mathcal{H}_{\bar{A},0}$ as the sector where the particle is within $\bar{A}$.

Now we compute the reduced density matrix of a state with respect to $\mathcal{A}$. Let $\rho = |\psi\rangle \langle \psi|$ be a general pure state for
\begin{align}
|\psi\rangle & \equiv \int dx \, \psi(x) |x\rangle = \int_{x \in A}dx\,  \psi(x) |x\rangle + \int_{x \in \bar{A}}dx\,  \psi(x) |x\rangle \\
& \equiv |\psi_A\rangle + |\psi_{\bar{A}}\rangle.
\end{align}
We compute the reduced density matrix $\rho_A$ using Eqn.\  \ref{eqn:RDM}, taking note of the decomposition in Eqn.\  \ref{eqn:single-particle-decomposition}.  First we project the density matrix $\rho$ into each of the two sectors, yielding $ |\psi_A\rangle\langle \psi_A|$ and $ |\psi_{\bar{A}} \rangle\langle \psi_{\bar{A}}|$.    Following Eqn.\  \ref{eqn:probabilities}, define
\begin{align}
p_0  &\equiv \Tr( |\psi_A\rangle\langle \psi_A|) = \langle \psi_A | \psi_A\rangle, \\
p_1  &\equiv \Tr( |\psi_{\bar{A}} \rangle\langle \psi_{\bar{A}}|) = \langle \psi_{\bar{A}} | \psi_{\bar{A}}\rangle,
\end{align}
and

\begin{align}
\rho_0 & = \frac{1}{p_0}  |\psi_{\bar{A}} \rangle\langle \psi_{\bar{A}}| ,\\
\rho_1 & = \frac{1}{p_1}|\psi_A\rangle\langle \psi_A|.
\end{align}
Finally, plugging these into Equations \ref{eqn:RDM} and \ref{eqn:algebraic-EE}, we obtain
\begin{equation} \label{singleEE}
S(\rho,\mathcal{A}) = -p_0 \log(p_0) - p_1 \log(p_i),
\end{equation}
and we find the entanglement entropy has a contribution only from the classical term.  This classical piece is the Shannon entropy associated to the probabilities of the single particle appearing in $A$ or $\bar{A}$.  In the multi-particle case, we will see that there is generically a quantum piece as well.

\subsection{General target spaces}

Nothing in our construction relied on properties of the simple target line $\mathbb{R}$. Indeed, we may consider a particle moving on some general $d$-dimensional target space $T$, with coordinates $\vec{x}$. The Hilbert space is given by $L^2(T)$ and admits the same decomposition $\mc{H}=\mc{V}_{A}\oplus \mc{V}_{\bar{A}}$ where $A \cup \bar{A} = T$. We can take $A$ to as complicated a region as we would like.  We define the relevant subalgebra as 

\begin{equation}
    \mc{A}= \bigg \langle \{ \ket{\vec{x}}\bra{\vec{x}'} : \vec{x},\vec{x}'\in A \} \cup \mathds{1}_{\mc{H}} \bigg \rangle,
\end{equation}
which will again have non-trivial center.

All subsequent steps follow through straightforwardly. 

\subsection{Multiple indistinguishable particles}

We consider now the general set-up of $N$ particles propagating on a general target space $T$, for instance $T = \mathbb{R}^d$.   A large literature exists on the entanglement of identical particles \cite{Dalton2017}, including e.g.\ an algebraic approach in \cite{Balachandran2013}.  However, here we will study the entanglement with respect to partitions of $T$, \textit{not} the set of particles. 

\subsubsection{Bosons}  \label{sec:bosons}

We first study bosons. Denoting the single particle Hilbert space by $\mc{H}=L^2(T)$, the physical Hilbert space is the symmetric quotient

\begin{equation}
    \mc{H}_{N} \equiv \Sym(\mathcal{H}^{\otimes N}) \equiv \frac{\mc{H}^{\otimes N}}{S_N}
\end{equation}
where the $S_N$ quotient arises from the indistinguishability of the $N$ particles.   That is, $\mc{H}_{N}$ consists of permutation-symmetric wavefunctions $\psi(\vec{x}_1,...,\vec{x}_N)$.

Given a partition of the target space $T$ into complementary regions $A,\bar{A}$, we want to associate a sub-algebra of observables $\mathcal{A} \subset L(\mathcal{H}_{N})$ to $A$.

We propose the following algebra,
\begin{equation} \label{eqn:multipaticle-algebra}
    \mc{A} \equiv \bigg \langle \{ P_{S_N} \left( \ket{\vec{x}}_1 \bra{\vec{x}'}_1\otimes \mathds{1}_2 \otimes ... \otimes \mathds{1}_N  \right) P_{S_N} : \vec{x},\vec{x}' \in A \} \cup \mathds{1}_{\mc{H}_{N}} \bigg \rangle ,
\end{equation}
where $P_{S_N}$ is the projection onto the symmetric subspace of $\mc{H}^{\otimes N}$,
\begin{equation}
    P_{S_N} \equiv \frac{1}{N_!} \sum_{\sigma \in S_N} P_{\sigma},
\end{equation}
and where $P_\sigma$ permutes the subsystems according to the permutation $\sigma \in S_{N}$. The appearance of $P_{S_N}$ in Eqn.\ \ref{eqn:multipaticle-algebra} is crucial for generating all multi-particle operators. The subscripts on the kets are particle labels, denoting which copy of $\mc{H}$ within $\mc{H}^{\otimes N}$ the operator acts on. For instance, unpacking the notation for the case of $N=2$, we have
\begin{equation}
    P_{S_2} \left( \ket{\vec{x}}_1 \bra{\vec{x}'}_1 \otimes \mathds{1}_2 \right) P_{S_2}= \frac{1}{2!} \left ( \ket{\vec{x}}_{1}\bra{\vec{x}'}_1 \otimes \mathds{1}_2 + \mathds{1}_1 \otimes \ket{\vec{x}}_2 \bra{\vec{x}'}_2 \right) .
\end{equation}

To motivate this algebra operationally, note that in ordinary quantum mechanics, if an external apparatus $\mc{X}$ situated in region $A$ were coupled to the system of identical particles $\mc{H}_N$ in a way that respected permutation symmetry and particle-number conservation, the apparatus $\mc{X}$ could only be coupled with a Hamiltonian of the form 
\begin{align} \label{eqn:apparatus-interaction}
H_{int} = \sum_i O_i^X \otimes O_i^A
\end{align}
for operators $O_i^X \in L(\mc{X})$ and $O_i^A \in \mc{A}$.  (The fact that interactions must take this form may be more obvious from the form of the algebra in Eqn.\ \ref{eqn:multiparticle-decomposition}.) If observers on $A$ and $\bar{A}$ are allowed to perform operations only using such apparatuses, the amount of entanglement distillable through local operations and classical communication (LOCC) will be equal to the (quantum piece of the) entanglement entropy with respect to $\mc{A}$.  This operational interpretation follows as a corollary to the discussions in \cite{schuch2004nonlocal,Verstraete2016,soni2016aspects}.

To better understand the above algebra, note that we can decompose
\begin{align} \label{eqn:multiparticle-decomposition}
    \mc{H}_{N} & \equiv \frac{\mc{H}^{\otimes N}}{S_N} = \frac{(\mc{V}_A \oplus \mc{V}_{\bar{A}})^{\otimes N}}{S_N}  \nonumber \\
& = \bigoplus^{N}_{k=0} \frac{\mc{V}_A ^{\otimes k}}{S_k} \otimes \frac{\mc{V}_{\bar{A}} ^{\otimes N-k k}}{S_{N-k}}.
\end{align}
where we define $\mc{V}_{A}^0=\mc{V}_{\bar{A}}^0=\mathbb{C}$. 
The sectors indexed by $k$ in the sum correspond to states with $k$ particles in $A$ and $N-k$ particles in $\bar{A}$.  It turns out that when our algebra $\mathcal{A}$ of Eqn.\  \ref{eqn:multipaticle-algebra} above is decomposed in the general way of Eqn.\  \ref{eqn:decomposition}, we obtain precisely the above decomposition.  That is, schematically, we have
\begin{align} \label{eqn:multiparticle-algebra-decomposition}
\mathcal{A} = \bigoplus^{N}_{k=0} L\left(\frac{\mc{V}_A ^{\otimes k}}{S_{k}}\right) \otimes \mathds{1}_{\Sym(\mc{V}_{\bar{A}}^{\otimes N-k})}.
\end{align}
To justify the above using the definition in Eqn.\  \ref{eqn:multipaticle-algebra}, see Appendix A.  The above demonstrates the algebra decomposes according to the particle number ``superselection'' sectors reviewed in \cite{Dalton2017}.

We may now write down the reduced density matrix.  Let $\Pi_k$ be the projector onto the $k$'th sector in the decomposition of Eqn.\  \ref{eqn:multiparticle-decomposition}; for an explicit expression, see Appendix A.  Following the definition for the reduced density matrix in Eqn.\  \ref{eqn:RDM}, we have 
\begin{equation}
    \rho_A = \sum _{k=0}^N p_k \Tr_{\textrm{Sym}(\mc{V}_{\bar{A}}^{\otimes N-k})}\rho_k \otimes \frac{\mathds{1}_{\textrm{Sym}(\mc{V}_{\bar{A}}^{\otimes N-k})}}{\textrm{dim}(\textrm{Sym}(\mc{V}_{\bar{A}}^{\otimes N-k}))}
\end{equation}
where

\begin{equation}
    p_k \equiv  \Tr_{\mc{H}_{N}} \left(  \Pi_k \rho \Pi_k \right)
\end{equation}
and
\begin{equation}
\rho_k \equiv \frac{1}{p_k} \left( \Pi_k\rho \Pi_k \right) 
\end{equation}

In particular, for a pure state, we have
 \begin{equation}
    p_k = \binom{N}{k} \int_{A} dx_{1} ... dx_{k} \int_{\bar{A}} dx_{k+1}...dx_{N} |\psi(x_{1},...,x_{N})|^2,
\end{equation}
which is the probability of finding $k$ particles in $A$.  

We can immediately compute the classical part of the entanglement entropy, $\sum_k -p_k \log(p_k)$, corresponding to the Shannon entropy for finding varying numbers of particles in $A$ and $\bar{A}$.  Meanwhile, unlike for the case of a single particle, here the blocks $\rho_k$ of the density matrix are generically entangled between $A$ and $\bar{A}$, so the quantum term of the entanglement entropy is nonzero.

\subsubsection{Fermions}
All the machinery we have built generalizes quite simply to fermions. In that case, we need to consider the algebra 

\begin{equation}
    \mc{A}_F =   \bigg \langle \{ P_{\text{Asym}_N} \left( \ket{\vec{x}}_{1}\bra{\vec{x}'}_1\otimes \mathds{1}_2 \otimes ... \otimes \mathds{1}_N  \right) P_{\text{Asym}_{N}} : \vec{x},\vec{x}' \in A \} \cup \mathds{1}_{\mc{H}_{N}} \bigg \rangle  
\end{equation}
where $P_{\text{Asym}_N}$ is the projector onto the anti-symmetric subspace of $\mc{H}^{\otimes N}$, defined via 

\begin{equation}
    P_{\text{asym}_N} = \frac{1}{N!} \sum_{\sigma \in S_{N}} (-1)^\sigma P_{\sigma}.
\end{equation}

\section{Comparison with embedding into second-quantized theory} \label{sec:Fock}
To define entanglement in a first-quantized theory of many particles, rather than use an algebraic definition, we could also translate to the second-quantized picture where a natural tensor factorization  does exist.  Here, we consider the latter approach and confirm that it agrees with the calculations of the previous section.  

In the second-quantized approach, we consider the $N$-particle Hilbert space 
\begin{equation}
    \mc{H}_{N} \equiv \Sym(\mathcal{H}^{\otimes N})
\end{equation}
as a subspace of the Fock space
\begin{equation}
    \mc{H}_F \equiv \bigoplus_{N=0}^\infty \Sym(\mathcal{H}^{\otimes N}).
\end{equation}
Let us re-phrase the familiar process of second quantization as the process whereby, given a basis of the single-particle Hilbert space $\mathcal{H}$, we induce a tensor factorization of the Fock space $\mathcal{H}_F$.  For instance, choosing the position basis of $\mathcal{H}$, we write the Fock space as
\begin{align} \label{eqn:Fock-factorization}
\mc{H}_F = \bigotimes_{\vec{x}} \mathcal{H}_{\vec{x}}
\end{align}
where $\mathcal{H}_{\vec{x}} = \textrm{span}\{|0\rangle_{\vec{x}},|1\rangle_{\vec{x}},...\}$ is the countably infinite-dimensional Hilbert space whose basis states $|n\rangle_{\vec{x}}$ indicate $n$ particles occupying position $\vec{x}$.\footnote{The above tensor product  is purely formal; it's a continously indexed tensor product.  However, if we chose a countable basis for the single-particle Hilbert space $\mathcal{H}$, rather than the naive basis of position kets, the above tensor product would be countably indexed, so that it could be made rigorous.}

For concreteness, using the factorization of Eqn.\  \ref{eqn:Fock-factorization}, the zero-particle state in the Fock space looks like $\bigotimes_{\vec{x}} |0\rangle_{\vec{x}} \in \mc{H}_F$, and the single-particle state $|\vec{y}\rangle \in \mathcal{H}_1$ embeds into the Fock space as $\left(\bigotimes_{\vec{x} \neq \vec{y}} |0\rangle_{\vec{x}} \right) \otimes |1\rangle_{\vec{y}} \in \mc{H}_F$.

More generally, by defining raising and lower operators $a^{\dagger}_{\vec{x}}, a_{\vec{x}}$ for each factor $\mc{H}_{\vec{x}}$ such that $\ket{n}_{\vec{x}}= \frac{a^{\dagger}_{\vec{x}}}{\sqrt{n!}}\ket{0}_{\vec{x}}$, we can neatly rewrite the embedding of the state 

\begin{equation}
\ket{\psi}= \int d\vec{x}_{1}...d\vec{x}_{N}\psi(\vec{x}_{1},....,\vec{x}_{N})\ket{x_1}\otimes....\otimes \ket{x_N} \in \mc{H}_N 
\end{equation}

as a state in the Fock space 

\begin{equation}
    \ket{\psi}_{F}=\int d\vec{x}_{1}...d\vec{x}_{N}\psi(\vec{x}_{1},....,\vec{x}_{N}) a^{\dagger}_{\vec{x}_1}...a^{\dagger}_{\vec{x}_N} \left( \otimes_{\vec{x}}\ket{0}_{\vec{x}} \right) \in \mc{H}_{F}
\end{equation}

The ``target space'' of the first-quantized theory thus becomes the base space of the second-quantized theory.  The partition of the target space in the first-quantized theory becomes an ordinary partition of the base space for the second-quantized theory.  Given a region $A$, we want to check that the algebraic entanglement entropy $S(\rho,\mathcal{A})$ of a pure state $\rho$ living on $\mc{H}_{N}$ matches the ordinary entanglement entropy of  $\rho$ when viewed as a state on $\mc{H}_F$.  

For a region $A$, we decompose the single-particle Hilbert space as
\begin{align}
\mathcal{H} = \mathcal{V}_A \oplus \mathcal{V}_{\bar{A}}.
\end{align}
Then we can define a Fock space for $\mc{V}_A$,
\begin{align}
 (\mc{V}_A)_F & \equiv \bigoplus_{N=0}^\infty \Sym(\mathcal{V}_A^{\otimes N}) \\
& =  \bigotimes_{\vec{x} \in A } \mathcal{H}_{\vec{x}}, \nonumber
\end{align}
and likewise for $\bar{A}$.  

The entire Fock space therefore factorizes as 
\begin{align} \label{eqn:Fock-factorization-A}
\mc{H}_F  & = (\mc{V}_A)_F  \otimes (\mc{V}_{\bar{A}})_F \\
& = \left( \bigoplus_{N=0}^\infty \Sym(\mathcal{V}_A^{\otimes N}) \right) \otimes \left( \bigoplus_{N=0}^\infty \Sym(\mathcal{V}_{\bar{A}}^{\otimes N}) \right) \\
& = \ \bigoplus_{N,M = 0}^\infty \Sym(\mathcal{V}_A^{\otimes N})\otimes \Sym(\mathcal{V}_{\bar{A}}^{\otimes M}) \\
& \supset  \bigoplus^{N}_{k=0} \Sym(\mc{V}_A^{\otimes k}) \otimes  \Sym(\mc{V}_{\bar{A}}^{\otimes N-k}) = \mathcal{H}_N.
\end{align}
where the last line uses the decomposition of Eqn.\  \ref{eqn:multiparticle-decomposition}.  Thus we embed $\mc{H}_N \subset \mc{H}_F$ in a way neatly compatible with the factorization into $A$, $\bar{A}$.

Combining the above embedding with Eqn.\  \ref{eqn:multiparticle-algebra-decomposition}, which illustrates the structure of the algebra $\mathcal{A}$ in the first-quantized picture, and recalling the definition of $S(\rho,\mathcal{A})$ in either Eqn.\  \ref{eqn:algebraic-EE} or \ref{eqn:algebraic-EE-ext}, we conclude that 
\begin{align}
S(\rho,\mathcal{A}) = S(\rho_F)
\end{align}
where $\rho_F$ indicates the state $\rho$ embedded in the Fock space $\mc{H}_F$. $S(\rho_F)$ is the ordinary von Neumann entanglement entropy with respect to the factorization of Eqn.\  \ref{eqn:Fock-factorization-A}, i.e.\ 
\begin{equation}
    S(\rho_F)= \Tr_{A} \left( \rho_A \log \rho_A \right).
\end{equation}

\section{Competing notions of locality in relativistic field theories} \label{sec:axVSphix}

In Section \ref{sec:Fock} we saw that the algebraic entanglement entropy in the first-quantized setting agrees with the ordinary entanglement entropy in the second-quantized setting.  However, we must take care with relativistic field theories, where we find two competing notions of locality. While one appears quite natural from the first quantized perspective, the other serves as the standard in most QFT calculations of entanglement entropy.

Consider the free scalar field in $d+1$ dimensions.  We will discuss two alternative factorizations of the Hilbert space, given a spatial partition.  Similar discussion appears already in \cite{Piazza2007}.  Afterward, we return to the subject of algebraic entanglement entropy.

Let us start by reviewing the ``ordinary'' spatial factorization of a quantum field theory, ignoring subtleties associated to the continuum \cite{witten2018aps}.  While the content may be familiar, we must be explicit to avoid confusion between the alternative factorizations.

The Hilbert space formally factorizes as 
\begin{align} \label{eqn:QFT-factorization}
\mc{H}_{QFT} = \bigotimes_{\vec{x} \in \mathbb{R}^d} \mc{P}_{\vec{x}}
\end{align}
where 
\begin{align}
\mc{P}_{\vec{x}} \equiv \textrm{span}\{|\phi\rangle_{\vec{x}} : \phi \in \mathbb{R} \} \cong L^2(\mathbb{R})
\end{align}
is the Hilbert space associated to the field degree of freedom living at base point $\vec{x}$.  This is the ordinary tensor factorization of a field theory.  When free field theory is viewed as a collection of coupled harmonic oscillators, $\mc{P}_{\vec{x}}$ is the Hilbert space of the harmonic oscillator ''living" at $\vec{x}$. 

The field operator $\hat{\phi}(\vec{x})$ living at a point $\vec{x}$ is local to the tensor factor $\mc{P}_{\vec{x}}$, and it acts on states $ \phi |\phi\rangle_{\vec{x}} \in \mc{P}_{\vec{x}}$ as
\begin{align}
\hat{\phi}(\vec{x}) |\phi\rangle_{\vec{x}} = \phi |\phi\rangle_{\vec{x}}.
\end{align}

Given a field configuration $\phi : \mathbb{R}^d \to \mathbb{R}$ denoted $\phi(x)$, one can then define a field ket $|\phi\rangle \in \mc{H}_{QFT}$ as the simultaneous eigenstate of the field operators $\hat{\phi}(\vec{x})$ with respective eigenvalues $\phi(x)$.  That is,
\begin{align}
|\phi\rangle \equiv \bigotimes_{\vec{x} \in \mathbb{R}^d} |\phi(x)\rangle_{\vec{x}}.
\end{align}
Finally, the wavefunctional $\Psi[\phi]$ expands an arbitrary state in $\mc{H}_{QFT}$ in terms of field kets $|\phi\rangle$.  Given a region $A \subset \mathbb{R}^d$ and complementary region $\bar{A}$, we obtain a bipartite factorization
\begin{align}
\mc{H}_{QFT} = \mc{H}_{QFT, A} \otimes \mc{H}_{QFT, \bar{A}}
\end{align}
where
\begin{align}
\mc{H}_{QFT, A} = \bigotimes_{\vec{x} \in A} \mc{P}_{\vec{x}}
\end{align}
and likewise for $ \mc{H}_{QFT, \bar{A}}$.

We will call this factorization of the Hilbert space the ``ordinary'' or ``field-based'' factorization.  It is the usual factorization used to define entanglement in relativistic field theories, wherein the vacuum exhibits an area law divergence in entanglement entropy.  (Again, for a continuum field theory, this ordinary factorization does not actually exist as a tensor product \cite{witten2018aps}.)

Meanwhile, we also have a ``Fock-based'' factorization of the Hilbert space, akin to the factorization expressed in Eqn.\  \ref{eqn:Fock-factorization}.  We utilize the Fock struture of the free theory,
\begin{align}
\mc{H}_{QFT} \cong \mc{H}_F \equiv \bigoplus_{N=0}^\infty \Sym(\mathcal{H}^{\otimes N}),
\end{align}
where $\mc{H}$ is the single-particle Hilbert space.  How do we identify the two Hilbert spaces above?  We can use the momentum basis  $\mc{H} = \textrm{span}\{|\vec{p}\rangle\}$ for the single-particle space.  Let $a_{\vec{p}}, a_{\vec{p}}^\dagger$ be the ladder operators that raise/lower the occupancy of the $|\vec{p}\rangle$ state in the Fock space.  If we identify $|\vec{p}\rangle \in \mc{H}$ with the single-particle momentum eigenstate in the free field theory, $|\vec{p}\rangle \in \mc{H}_{QFT}$, then the ladder operators are related to the field operators in the usual way (taking $d+1=3+1$ for simplicity)
\begin{align}
\hat{\phi}(\vec{x}) & = \int \frac{d^3\vec{p}}{(2\pi)^3} \frac{1}{\sqrt{2E_{\vec{p}}}} \left(a_{\vec{p}} e^{i\vec{p}\cdot \vec{x}} + a_{\vec{p}}^\dagger e^{-i\vec{p}\cdot \vec{x}} \right), \\
E_{\vec{p}} & \equiv \sqrt{\vec{p}^2 + m^2},
\end{align}
using the normalization conventions of \cite{Peskin1995}.

If we instead choose the position basis for the single-particle Hilbert space $\mc{H} = \textrm{span}\{|\vec{x}\rangle\}$, with the momentum and position basis related by the ordinary Fourier transform, we can define the coresponding ladder operators that raise/lower the occupancy of the $| \vec{x}\rangle$ state in the Fock space.  These are given by 
\begin{align}
a_{\vec{x}} & \equiv \int \frac{d^3\vec{p}}{(2\pi)^3} e^{i\vec{p}\cdot \vec{x}} a_{\vec{p}}, \\
a_{\vec{x}}^\dagger & \equiv \int \frac{d^3}{(2\pi)^3} \vec{p} e^{-i\vec{p}\cdot \vec{x}} a_{\vec{p}}^\dagger,
\end{align}
and these ladder operators are local to the factors of the tensor factorization in Eqn.\  \ref{eqn:Fock-factorization},
\begin{align}
\mc{H}_F = \bigotimes_{\vec{x}} \mathcal{H}_{\vec{x}},
\end{align}
where the local Hilbert spaces $\mathcal{H}_{\vec{x}} = \textrm{span}\{|0\rangle_{\vec{x}},|1\rangle_{\vec{x}},...\}$ have basis states$|n\rangle_{\vec{x}}$ that count the number of particles occupying single-particle state $\vec{x}$.

This defines the ``Fock-based'' factorization referred to above.  The ladder operators $a_{\vec{x}}\dagger, a_{\vec{x}}$ raise and lower the particle number of the free theory.  In this factorization, the vacuum is just the zero-particle state $\bigotimes_{\vec{x}} |0\rangle_{\vec{x}}$.  Note this is a product state!  The vacuum is unentangled with respect to the Fock factorization.  Clearly, the Fock-based factorization differs from the ordinary field-based factorization.

To sharpen the distinction between the factorizations, let us define ladder operators $\alpha_{\vec{x}}, \alpha^\dagger_{\vec{x}}$ associated to the ``harmonic oscillator'' Hilbert space $\mc{P}_{\vec{x}}$,  the local degrees of freedom in the ordinary tensor factorization.  That is, take
\begin{align}
\alpha_{\vec{x}} & = \frac{1}{\sqrt{2}} ( \hat{\phi}(x) + i\hat{\pi}(\vec{x}) ),  \\
\alpha^\dagger_{\vec{x}} & =\frac{1}{\sqrt{2}}( \hat{\phi}(x) - i\hat{\pi}(\vec{x}) ) \nonumber 
\end{align} 
were $\hat{\pi}(\vec{x})$ is the canonical conjugate of the field operator $\hat{\phi}(\vec{x})$, acting as $-i \frac{\delta}{\delta \phi(\vec{x})}$ on the wavefunctional.  Note these are $\textit{not}$ the same as the ladder operators $a_{\vec{x}}\dagger, a_{\vec{x}}$ associated to the Fock-based factorization.  

Any operator local to $\vec{x}$ in the ordinary factorization should commute with $\phi(\vec{y})$ for all $\vec{y} \neq \vec{x}$, whereas 
\begin{align}
[a_{\vec{x}}, \hat{\phi}(\vec{y})] \propto  K(\vec{x},\vec{y}),
\end{align}
where $K(\vec{x},\vec{y})$ is the convolution kernel 
\begin{align}
K(\vec{x},\vec{y}) \equiv \int \frac{d^3\vec{p}}{(2\pi)^3} e^{i\vec{p}\cdot(\vec{x}-\vec{y})} \frac{1}{\sqrt{2E_{\vec{p}}}},
\end{align}
emphasizing that the operators $a_{\vec{x}}\dagger, a_{\vec{x}}$ local in the Fock-based factorization are slightly \textit{non-local} in the ordinary field-based factorization.

In one sense, the two factorizations are ``close,'' because the kernel $K(x,y)$ is peaked near  $\vec{x} \sim \vec{y}$.  Thus an operator local to a region $A$ in the Fock-based factorization will be well-approximated by an operator local to a sufficiently larger $B \supset A$ in the ordinary factorization.

In another sense, the alternatives yield drastically different entanglement entropies: the vacuum is unentangled in the Fock-based factorization, while it exhibits diverging entanglement in the ordinary factorization.  It turns out that multi-particle excited states yield a middle ground: if the wavefunctions of the particles are sufficiently spread, the two factorizations will yield approximately equal entanglement entropies, up to a correction which is precisely the vacuum entanglement.  

Which factorization is ``correct''?  Of course they merely constitute different choices.   If we want to leverage the operational interpretation of entanglement entropy, we must ask which algebra of observables is available to an observer who ``has access to region $A$'' ? We will not further pursue this question, but a point in favor of the ordinary factorization is that the Hamiltonian is truly local with respect to this factorization.  Moreover, only in the ordinary factorization is there a strict lightcone, i.e.\ exact commutation of spacelike-separated Heisenberg operators. 

\section{Computation of entanglement entropy for finite-particle states: ``Fock''- vs.\ ``field''-based factorization}\label{computation}

Our algebraic setup calculates the entanglement entropy relative to the Fock-based tensor product factorization of the QFT. In this section, we show there is a sense in which the entanglement entropy of a multi-particle state decomposes into the universal, divergent area-law piece and an additive contribution we can associate to the particles' wavefunction. The extra entanglement due to the excitations has been called the ``excess of entanglement'' above the vacuum \cite{pizorn2012universality,castro2018entanglement}. We will address the simple case of single-particle excitations, but the account of finitely multi-particle excitations is similar.

Our calculation of the entanglement entropy of single-particle excitations (with respect to the ordinary tensor factorization) has precedent in the related calculations of \cite{pizorn2012universality,castro2018entanglement}.  However, those arguments only apply to momentum eigenstates, rather than to excited states with more general wavefunctions.  The argument sketched here has a different scope.

In the Fock basis, we can describe a single particle state as 

\begin{equation} \label{axstate}
     \ket{\psi}= \int d^{3}x \psi(x) a^{\dagger}_{\vec{x}} \ket{0} = \int \frac{d^{3}p}{(2\pi )^3} \tilde{\psi}(p) a^{\dagger}_{\vec{p}} \ket{0}
\end{equation}

The entanglement entropy for a spatial subgion $A$, can be immediately computed as 

\begin{equation}
  H(\{p_{A}^{(a_x)},1-p_{A}^{(a_x)}\}) = -p_{A}^{(a_x)} \log(p_{A}^{(a_x)}) - (1-p_{A}^{(a_x)})\log(1-p_{A}^{(a_x)})
\end{equation}
with 
\begin{equation}
   p_{A}^{(a_x)} = \int_{A} d^{3}x |\psi(x)|^2
\end{equation}

We wish now wish to compare this to the entanglement entropy computed relative to the field-based factorization. First, let us rewrite the state $\ket{\psi}$ as
\begin{equation} \label{phixstate}
     \ket{\psi}= \int d^{3}x f(x) \hat{\phi}(\vec{x}) \ket{0} = \int \frac{d^{3}p}{(2\pi )^3} \frac{\tilde{f}(p)}{\sqrt{2E_{\vec{p}}}} a^{\dagger}_{\vec{p}} \ket{0}
\end{equation}

Eqn.\ \ref{axstate} therefore identifies $\tilde{\psi}(p)=\frac{\tilde{f}(p)}{\sqrt{2E_{\vec{p}}}}$, or alternatively, in position space $\psi(\vec{x})=\int{d\vec{y}} K(\vec{x},\vec{y})f(y)$. 

Below, we give a proof (on the lattice) that we may well approximate the entanglement entropy as 
\begin{align} \label{eqn:approxEE1part}
S(\rho_A) \approx S_0 + H(\{ p_A,1-p_A \})
\end{align}
where $S_0$ is the entanglement  of the vacuum and $H(\{p_A,1-p_A\}) = -p_A \log(p_A) - (1-p_A)\log(1-p_A)$ is the Shannon entropy of the classical probability distribution, but now with $p_A = \int_{A} d^{3}x |f(\vec{x})|^2$. 

Before delving into the mechanics of the proof, we stress we may meaningfully compare the Shannon entropies $H(\{p_A,1-p_A\})$ and $ H(\{p_{A}^{(a_x)},1-p_{A}^{(a_x)}\})$. When the kernel $K(\vec{x},\vec{y})$ is narrowly peaked near $\vec{x} \sim \vec{y}$, and the regions $A$ are taken sufficiently large, these quantities are in fact close (at least on the lattice). In fact, 
\begin{align}
H(\{p_{A}^{(a_x)},1-p_{A}^{(a_x)}\}) \to H(\{p_A,1-p_A\}) 
\end{align}
precisely in the limit described in Section \ref{sec:excitation-EE-proof} below, as the wavefunctions are spread over large regions.

\subsection{Proof} \label{sec:excitation-EE-proof}
Consider a free, massive scalar field on a finite square lattice in $d$ spatial dimensions.  The discretized field theory Hamiltonian is that of coupled harmonic oscillators,
\begin{align}
H = \sum_{x_i} \phi(x_i)^2 + \pi(x_i)^2 + \sum_{\langle x_i,x_j \rangle} m^2 (\phi(x_i) - \phi(x_j))^2,
\end{align}
for fields $\phi(x_i)$ at site $i$ and conjugate momenta $\pi(x_i)$. We consider the single-particle excitation
\begin{align}
|\psi\rangle = \sum_{x_i} f(x_i) \phi(x_i) |\Omega\rangle,
\end{align}
not necessarily an energy or momentum eigenstate, where $|\Omega\rangle$ is the vacuum, and $f(x_i)$ is some ``wavefunction'' of the discrete positions $x_i$, normalized so that the overall state is normalized.\footnote{However, note the norm of $|\psi\rangle$ is not given by $\sum_i |f(x_i)|^2$, because the states $\phi(x_i)|\Omega\rangle$ are not orthogonal for distinct $i$.} Partition the lattice into complementary, contiguous regions $A, \bar{A}$, and consider the reduced state $\rho_A$.  We want to show that 
\begin{align} \label{eqn:excess-EE}
S(\rho_A) \approx S_0 + H(\{p_A,1-p_A\})
\end{align}
where $S_0$ is the entanglement  of the vacuum, $H(\{p_A,1-p_A\}) = -p_A \log(p_A) - (1-p_A)\log(1-p_A)$ is the Shannon entropy of the classical probability distribution, and where
\begin{align}
p_A \equiv \sum_{x_i \in A} |f(x_i)|^2
\end{align}
is essentially the probability of finding the particle in $A$ (at least for large $A$, due to subtleties about measuring particle position in this context).  Eqn.\ \ref{eqn:excess-EE} will hold with small error when the system has large volume and the wavefunction $f_i$ is not too concentrated around the boundary of $A, \bar{A}$.  To be more precise, let $X_R \subset A$ be the sub-region of $A$ consisting of sites at a distance larger than $R$ lattice units from the boundary $\partial A$, and let $B_R = A \backslash X_R$ be the buffer region between $X$ and $A$.  We can quantify the amount of the wavefunction $f(x_i)$ concentrated in the buffer region as 
\begin{align}
 p_{B_R} \equiv \sum_{x_i \in B_R = A \backslash X_R} |f(x_i)|^2.
\end{align}
We will prove that for $p_{B_R}$ sufficiently small for a choice of buffer size $R$ sufficiently large, and for total lattice volume sufficiently large, Eqn.\ \ref{eqn:excess-EE} holds to arbitrarily good approximation.  That is, we show Eqn.\ \ref{eqn:excess-EE} holds exactly in the limit of a sequence of systems and wavefunctions where $|A|,|\bar{A}| \to \infty$, and $p_{B_R} \to 0$ for a choice of buffer sizes $R \to \infty$.  One could also prove the result with more fine-grained error analysis, but proving the simple limit will serve our illustration.  

Now we sketch the proof.

\textbf{Proof sketch.}  Divide the state into two terms
\begin{align}
|\psi\rangle&  =  \sum_{x_i \in A} f(x_i) \phi(x_i) |\Omega\rangle + \sum_{x_i \in \bar{A}} f(x_i) \phi(x_i) |\Omega\rangle \\
& \equiv |\psi_A\rangle + |\psi_{\bar{A}}\rangle.
\end{align} 
We can approximate the state instead as 
\begin{align}
|\psi\rangle & \approx |\widetilde{\psi}\rangle  \equiv \sum_{x_i \in X_R} f(x_i) \phi(x_i) |\Omega\rangle + \sum_{x_i \in \bar{A}} f(x_i) \phi(x_i) |\Omega\rangle \\
& \equiv |\psi_{X_R} \rangle + |\psi_{\bar{A}}\rangle.
\end{align} 
Then
\begin{align}
\langle \widetilde{\psi} | \psi \rangle = 1 - \langle \psi_{B_R} | \psi \rangle \to 1
\end{align}
in the given limit where $p_{B_R} \to 0$, so $ |\widetilde{\psi}\rangle $ approaches $|\psi\rangle$ in trace-distance.  Then the entanglement entropy of $ |\widetilde{\psi}\rangle $  approaches the entanglement entropy of  $|\psi\rangle$, using the continuity of the entanglement entropy with respect to trace distance \cite{Eisert2002}.  The continuity result of \cite{Eisert2002} requires the same assumptions as those discussed in Section \ref{sec:finiteness}, which the single-particle states here satisfy.\footnote{In a more detailed argument, some care must be taken with how the continuity bound depends on lattice size.}   Thus we can examine the entanglement entropy of $ |\widetilde{\psi}\rangle $ rather than  $|\psi\rangle$.   The reduced density matrix has four terms
\begin{align} \label{eqn:RDM-four-terms}
\Tr_{\bar{A}}(|\widetilde{\psi}\rangle \langle \widetilde{\psi} |) = \Tr_{\bar{A}}\left(|\psi_{B_R} \rangle \langle \psi_{B_R}|
+ |\psi_{B_R} \rangle \langle \psi_{\bar{A}}| + |\psi_{\bar{A}} \rangle \langle \psi_{B_R}| + |\psi_{\bar{A}} \rangle \langle \psi_{\bar{A}}| \right).
\end{align}
Let's start with the fourth term, call it $\sigma_A \equiv \Tr_{\bar{A}}\left(|\psi_{\bar{A}} \rangle \langle \psi_{\bar{A}}| \right)$.  Note that the connected correlation functions of local operators exponentially decay with distance in this massive free lattice theory.  Actually, we use the stronger fact that the mutual information $I(B_R : \bar{A})$ in the vacuum tends to zero as the size $R$ of the buffer region increases, which can be shown with the methods of \cite{casini2009entanglement}.  Then the connected correlation of \textit{any} bounded operators on $B_R$ and $\bar{A}$ must tend to zero for large $R$, using the fact that mutual information upper bounds connected correlations, by Pinsker's inequality.
Thus for any operator $O_A$ on $A$ with operator norm 1,  
\begin{align}
\Tr(\sigma O_A) & = \langle \Omega | O_A \left( \sum_{x_i \in \bar{A}} f(x_i) \phi(x_i) \right) ^2 | \Omega \rangle \\
& \to  \langle \Omega | O_A | \Omega \rangle \langle \Omega | \left( \sum_{x_i \in \bar{A}} f(x_i) \phi(x_i) \right) ^2 | \Omega \rangle \\
& \to \langle \Omega | O_A | \Omega \rangle (1-p_A),
\end{align}
where again all limits are taken as described above.\footnote{Because the operators $\phi(x_i)$ are not bounded, and the mutual information only upper bounds connected correlations of bounded operators, we cannot directly apply the upper bound as stated.  However, it turns out the action of $\phi(x_i)$ on the vacuum can be sufficiently well-approximated by the action of a bounded operator for our purpose.}
Thus the state $\sigma/(1-p_A)$ approaches the reduced state of the vacuum.  

Likewise, we have $\Tr_A \left(|\psi_{B_R} \rangle \langle \psi_{B_R}|\right)/p_A$ approaching the reduced state of the vacuum as well.  (Note the previous expression traces out $A$ rather than $\bar{A}$, but the entanglement entropy will be the same the same whether we trace out $A$ or $\bar{A}$.)

Now we analyze the second and third terms of Eqn.\ \ref{eqn:RDM-four-terms}.  These terms are actually equal by the Hermiticiy of $\phi$; call this term $\tau \equiv \Tr_{\bar{A}}( |\psi_{\bar{A}} \rangle \langle \psi_{B_R}|)$.  Similar to the above calculation, we have
\begin{align}
\Tr(\tau O_A) \to \langle \Omega | O_A | \Omega \rangle \langle \Omega |  \sum_{x_i \in \bar{A}} f(x_i) \phi(x_i) | \Omega \rangle = 0
\end{align}
for any fixed norm operator $O_A$ local to $A$, again using the fact that $I(B_R : \bar{A}) \to 0$ in the vacuum as $R$ increases.  The RHS is zero above simply because $\phi$ has zero vacuum expectation.  So $\tau \to 0$, and we can discard these terms.

Finally, the first and fourth terms of  Eqn.\ \ref{eqn:RDM-four-terms} tend to orthogonal operators, so the entropy of their sum is the average of their entropies, plus the Shannon entropy associated to their traces.  We conclude
\begin{align} 
S(\rho_A) \to  S_0 + H(\{p_A,1-p_A\})
\end{align}
in the given limit, as desired.


\section{Target space entanglement entropy for field theories}

Above we considered the spatial entanglement in first-quantized many-particle systems, alternatively interpreted as target space entanglement in a 0+1-dimensional theory.  Now we consider target space entanglement for a more general $d+1$-dimensional theory, i.e.\ with a higher-dimensional base space.  This quantity is more akin to what might be desired in worldsheet string theory, if one desires to partition the target spacetime.  However, worldsheet string theory will offer further complications due to re-parameterization invariance, as further discussed in Section \ref{sec:discussion}.

Consider a $d+1$-dimensional field theory defined on the base space $B$ (for example, $B\cong \mathbb{R}^d$), with its field $\phi$ taking values in the target space $T$.  Referring again to the ordinary base space factorization described in Eqn.\  \ref{eqn:QFT-factorization}, we have the formal expression
\begin{align} 
\mc{H}_{QFT} & = \bigotimes_{\vec{x} \in B} \mc{P}_{\vec{x}} \nonumber \\
\mc{P}_{\vec{x}} &  \equiv \textrm{span}\{|\phi\rangle_{\vec{x}} : \phi \in T \} \cong L^2(T). \nonumber
\end{align}
Consider a partition of target space $T$ into complementary regions $A, \bar{A} \subset T$. We associate the following algebra to the target space region $A$,

\begin{align}\label{eqn:QFT-target-algebra}
\mathcal{A} \equiv  \bigg \langle \{ \ket{\phi}_{\vec{x}}\bra{\phi'}_{\vec{x}} \otimes_{\vec{y}\neq \vec{x}} \mathds{1}_{\mc{P}_{\vec{y}}} : \phi,\phi'\in A, \forall \vec{x} \in B \} \cup \mathds{1}_{\mc{H}_{QFT}.}\bigg \rangle
\end{align}

Defining the projector $\Pi_A^{\vec{x}}$ on local Hilbert space $\mc{P}_{\vec{x}}$ at $\vec{x}$ as
\begin{align} \label{eqn:QFT-target-projector}
\Pi_A^{\vec{x}} \equiv \int_{\phi \in A} d\phi \, |\phi\rangle_{\vec{x}} \langle \phi |_{\vec{x}}.
\end{align}

we can immediately write down the center of $\mc{A}$

\begin{align}\label{eqn:QFT-target-center}
Z(\mathcal{A}) =  \bigg \langle \{ \Pi^{\vec{x}}_{A}\otimes_{\vec{y}\neq \vec{x}} \mathds{1}_{\mc{P}_{\vec{y}}}  :  \forall \vec{x} \} \cup \mathds{1}_{\mc{H}_{QFT}.}\bigg \rangle
\end{align}

These algebras resemble those of Section \ref{sec:bosons} because we may view our QFT as the first-quantized theory of many \textit{distinguishable} particles, labeled by $\vec{x}$, moving in target $T$ with coordinates $\phi$. 

With this definition in hand, we can calculate the target space entanglement entropy of general field theories defined on the lattice.  Unfortunately, the calculation does not appear straightforward. \footnote{We find this quite reminiscent of the early days of base space entanglement entropy, where analytical progress appeared similarly difficult. Both \cite{BombelliSorkin,Srednicki1993} ultimately resorted to numerical methods to discover the area law in the ground state of scalar field theory. What we need is a suitable analog of the powerful path integral replica trick.}  We settle for the simplest non-trivial example: a massive scalar ``field theory'' on two spatial lattice points, i.e.\ two coupled harmonic oscillators.

Consider two field degrees of freedom $\phi_1$ and $\phi_2$ at lattice points 1 and 2, with Hilbert space $\mc{H} = L^2(\mathbb{R}) \otimes L^2(\mathbb{R})$ and Hamiltonian
\begin{align} \label{eqn:2siteHamiltonian}
H = \frac{1}{2}(\pi_1^2 + \pi_2^2 + (\phi_1 - \phi_2)^2 + m^2(\phi_1^2 + \phi_2^2))
\end{align}
where $\pi_1, \pi_2$ are the conjugate momentum operators and $m$ the mass. The interaction term $(\phi_1 - \phi_2)^2$ comes from the lattice discretized spatial gradient of the field.

We choose to compute the entanglement entropy of the ground state,  partitioning the target space into the positive and negative half-lines, $A \equiv \{ \phi \in \mathbb{R} : \phi>0\}  \subset \mathbb{R}$. In this case, the algebra $\mc{A}$ of Eqn.\  \ref{eqn:QFT-target-algebra} has four sectors.  Defining the projectors $\Pi^1_A, \Pi^2_A, \Pi^1_{\bar{A}}, \Pi^2_{\bar{A}}$ as in Eqn.\  \ref{eqn:QFT-target-algebra}, the four sectors of $\mathcal{A}$ are simply the images of the four projectors $\Pi^1_A \Pi^2_A, \;  \Pi^1_A  \Pi^2_{\bar{A}}, \; \Pi^1_{\bar{A}}  \Pi^1_A , \; \Pi^1_{\bar{A}} \Pi^2_{\bar{A}}$.  The sector projected onto by $\Pi^1_A \Pi^2_A$ indicates ``the field on both lattice points takes values in $A$,'' while the sector projected onto by $\Pi^1_A  \Pi^2_{\bar{A}}$ indicates $\phi_1 \in A, \phi_2 \in \bar{A}$, and so on. Note there are four rather than three sectors because the lattice sites are distinguishable.   

The normalized ground state wavefunction for the Hamiltonian (\ref{eqn:2siteHamiltonian}) is given by
\begin{align} 
\psi(\phi_1,\phi_2) = \left( \frac{\omega_+ \omega_-}{\pi^2} \right)^{1/4} e^{-\frac{1}{2}( \omega_+ \phi_+^2  \, + \, \omega_- \phi_-)^2}
\end{align}
where $\phi_{\pm} = (\phi_1 \pm \phi_2)/\sqrt{2}$, $\omega_+ = m$, $\omega_- = \sqrt{m^2 + 2}$; see for instance the similar example in \cite{Srednicki1993}.


\begin{figure}[ht!]
\centering
\includegraphics[scale=0.4]{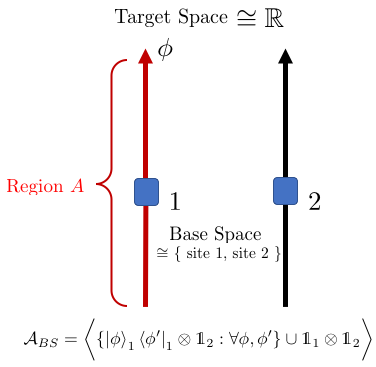}
\includegraphics[scale=0.4]{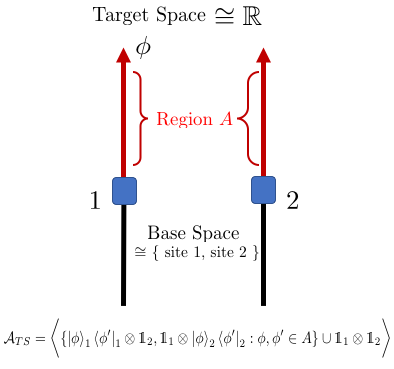} 
\caption{Base (Left) vs. Target (Right) Space Partition \& Associated Algebras for scalar field on two lattice sites. }
 \label{fig:two-site-regions}
\end{figure}

We consider the state $\rho = |\psi\rangle \langle \psi|$ projected separately onto the four sectors described above.  In the sector where $\phi_1, \phi_2 \in A$, the Hilbert space factorizes in a trivial way, as in the discussion surrounding Eqn.\  \ref{eqn:single-particle-decomposition} for the particle on a line. Hence the projection of the state onto this sector yields a product state, with no contribution to the quantum piece of the entanglement entropy.  The same holds for the sector associated to $\phi_1, \phi_2 \in \bar{A}$. The only contribution to the quantum part of the entanglement entropy thus comes from from the two sectors where $\phi_1, \phi_2$ are in different regions of the target space.  Since the groundstate is symmetric under the exchange of $\phi_1 \leftrightarrow \phi_2 $ the contribution in each such sector will be identical.  Thus we need only consider one sector, say the image of $\Pi^1_A  \Pi^2_{\bar{A}}$.  

The sector factorizes as $\mc{V}^1_A \otimes \mc{V}^2_{\bar{A}}$, where $\mc{V}^1_A \equiv \textrm{span}\{|\phi_1\rangle : \phi_1 \in A \}$ and $\mc{V}^2_{\bar{A}}  \equiv \textrm{span}\{|\phi_2\rangle : \phi_2 \in \bar{A} \}$.  We need to take the state projected on this sector, $\Pi^1_A  \Pi^2_{\bar{A}} |\psi\rangle \langle \psi | \Pi^1_A  \Pi^2_{\bar{A}}$, and trace out the second factor $\mc{V}^2_{\bar{A}}$.  We obtain the (non-normalized) density matrix $\sigma$ on $\mc{V}^1_A$ given by 
\begin{align} \label{eqn:two-site-RDM}
\sigma(x_1,y_1) = \int_{x_2 \in \bar{A}} dx_2 \, \psi(x_1,x_2) \psi(y_1,x_2)^*.
\end{align}
The integral above can be expressed in terms of error functions.  To calculate the entanglement entropy, it remains to diagonalize the above density matrix $\sigma$.   Returning attention to the full reduced state $\rho_{\mc{A}}$, the classical part of the entanglement entropy is then given by 
\begin{align}
S(\rho, \mc{A})_{classical} = H(\{p,p,1-p,1-p\})
\end{align}
where $H(\{\cdot\})$ is the classical (Shannon) entropy of the probability distribution, and $p = \Tr(\sigma)$, with $\sigma$ given above. Meanwhile, the quantum piece of the entanglement entropy is given by  
\begin{align} \label{eqn:two-site-EE-quantum}
S(\rho, \mc{A})_{quantum} = 2 p S\left(\sigma/\Tr(\sigma)\right).
\end{align}

\begin{figure}[ht]
\includegraphics[width=\textwidth]{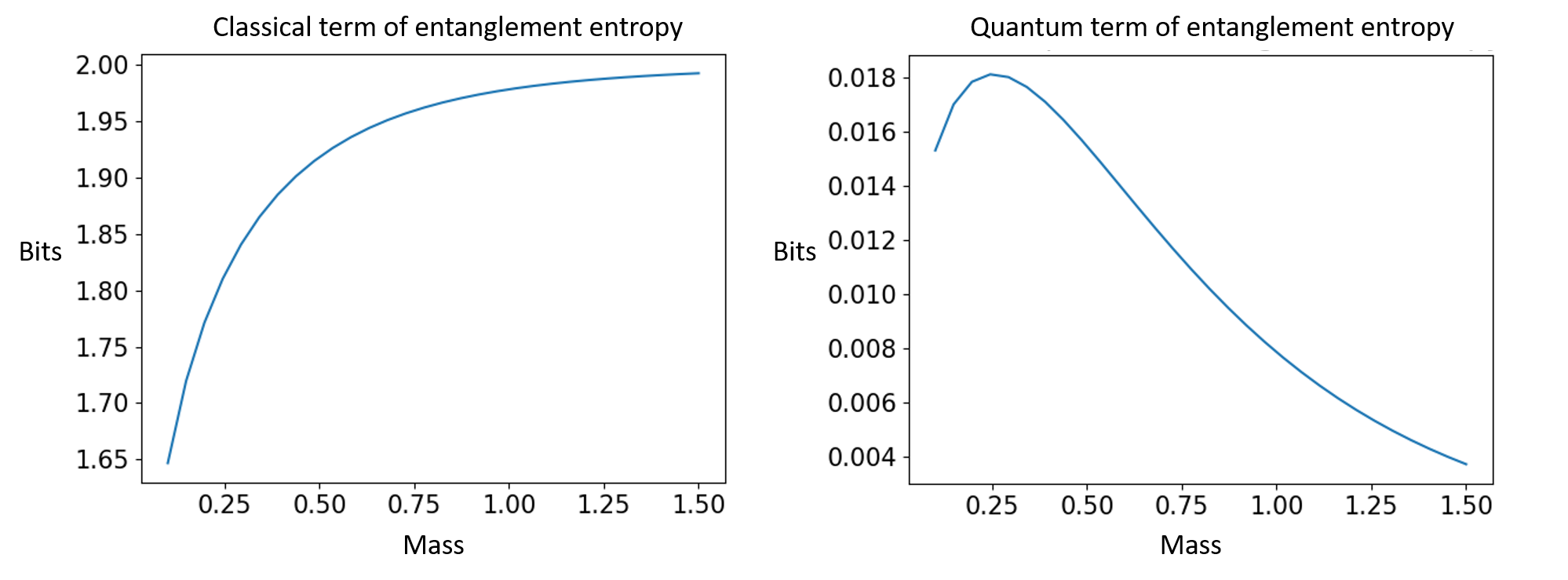}
\centering
\caption{Entanglement entropy with respect to a partition of target space, for two coupled oscillators governed by the Hamiltonian in Eqn.\ \ref{eqn:2siteHamiltonian}.}
 \label{fig:entropies}
\end{figure}

In lieu of an analytic method, we discretize the $x_1, y_1$ coordinates of Eqn.\ \ref{eqn:two-site-RDM} and numerically diagonalize the resulting finite matrix.  We ensure that the discretization is at sufficient resolution that the results converge when decreasing the spacing or increasing the total number of discretized points. Section \ref{sec:finiteness} guarantees convergence, the end result being finite. Ultimately we produce a numerical answer for the quantum and classical piece of the entanglement entropy of the ground state, as a function of the mass in the Hamiltonian.  The results are depicted in Figure \ref{fig:entropies}.  Numerical error due to discretization appears to be somewhat smaller than $10^{-3}$, but we do not include a rigorous analysis.

At high mass, the two harmonic oscillators approximately de-couple.  The wavefunction spreads equally between the four sectors, so that the classical piece of the entanglement entropy gives two bits.  Meanwhile, the quantum piece of the entanglement entropy tends to zero, because the only sectors that can contribute must have $\phi_1$ and $\phi_2$ in different regions $A$ and $\bar{A}$, and in these sectors, the wavefunction approximately factorizes due to the de-coupling of the oscillators.

Figure \ref{fig:entropies} also illustrates that the quantum term of the target space entanglement entropy is not monotonic with respect to the mass parameter. The non-monotonicity is associated with the fact that in Eqn.\ \ref{eqn:two-site-EE-quantum}, the first factor $p$ increases monotonically with mass, whereas the second factor $S(\sigma/\Tr(\sigma))$ decreases monotonically.

\section{Finiteness of the entanglement entropy} \label{sec:finiteness}

The entanglement entropies discussed in this paper involve infinite-dimensional Hilbert spaces and algebras.  In infinite dimensions, we must take care that density matrices and entropies remain well-defined.  Fortunately, we will see that most of the infinities present here are of a relatively tame variety.  

In this section, we will take more mathematical care, recalling for instance that the ``position eigenstate'' $|\psi\rangle$ is not a true state in the Hilbert space $L^2(\mathbb{R})$ as traditionally defined.  

The algebra associated to a region in multi-particle quantum mechanics (like Eqn.\ \ref{eqn:multipaticle-algebra}) is a finite direct sum of factors, where each factor is an infinite-dimensional ``Type $I$'' factor, according to the type theory of von Neumann algebras \cite{haag2012local}.  Type $I$ factors are algebras which are isomorphic to the full algebra of bounded operators on some Hilbert space.  The Type $I$ property of this algebra is therefore apparent from the schematic form of the algebra in Eqn.\ \ref{eqn:multiparticle-algebra-decomposition}.  

Similarly, an algebra associated to a region in the target space of a lattice field theory -- like the algebra in Eqn.\ \ref{eqn:QFT-target-algebra} on a finite lattice, or the algebra in Figure \ref{fig:two-site-regions} -- is also a direct sum of Type $I$ sectors, even when the target space is infinite-dimensional.  

For the general algebra decomposition of Eqn.\ \ref{eqn:decomposition} to make sense as written, it is indeed essential the algebra is a direct sum of Type $I$ factors. Otherwise the use of the tensor product there is incorrect.  

Even for these Type $I$ algebras, we must take care with the entanglement entropy.  The formula for the algebraic entanglement entropy in Eqn.\ $\ref{eqn:algebraic-EE}$ requires defining the von Neumann entropy of the partial trace of a pure state in a bipartite Hilbert space  $\mc{H}_1 \otimes \mc{H}_2$, where the factors $\mc{H}_1, \mc{H}_2$ may be countably infinite-dimensional. (The full algebraic entanglement entropy was then a sum of such entropies in each sector of the algebra.) We therefore focus on the question of ordinary von Neumann entanglement entropies of pure states in bipartite Hilbert spaces.  For any (mathematically legitimate, i.e.\ normalizable) state $|\psi\rangle \in \mc{H}_1 \otimes \mc{H}_2$, the partial trace $\rho = \Tr_2(|\psi\rangle\langle \psi|)$ can be taken using any (legitimate) orthonormal basis. The result will be a trace-class Hermitian operator $\rho$.  (To see that $\rho$ is trace class, we can take its trace in any orthonormal basis, and the resulting sum will be convergent by the normalizability of $|\psi\rangle$.) Recall that a trace-class Hermitian operator $\rho$ has an eigen-decomposition
\begin{align}
\rho = \sum_{i=1}^\infty \lambda_i |v_i\rangle \langle v_i |
\end{align}
for some countably infinite set of eigenvectors $\{|v_i\rangle\}$ and eigenvalues $\lambda_i$.  Thus we are in the position to define the entanglement entropy 
\begin{align}
S(\rho) \equiv \sum_{i=1}^\infty -\lambda_i \log(\lambda_i).
\end{align}
However, the above sum may be infinite, even though a normalized state $|\psi\rangle$ guarantees $\sum_i \lambda_i = 1$.  In fact, the set of states $|\psi\rangle$ with infinite entanglement entropy is dense in the total Hilbert space $\mc{H}_1 \otimes \mc{H}_2$, so in some sense the divergence is generic.  

Yet, for states of interest, the sum is often finite.  For instance, a finite energy condition may imply finiteness. The authors of \cite{Eisert2002} prove that, for any non-interacting Hamiltonian $H = H_1 \otimes \mathds{1}_2 + \mathds{1}_1 \otimes H_2$ on  $\mc{H}_1 \otimes \mc{H}_2$ with discrete spectrum such that $\Tr(e^{-\beta H})$ is finite for all $\beta > 0$, any state $|\psi\rangle \in \mc{H}$ that has finite expected energy $\langle \psi | H | \psi \rangle < \infty$ with respect to this Hamiltonian will have finite entanglement entropy.  Note the state $|\psi \rangle$ may have nonzero overlap with energy eigenstates of arbitrarily high energy; as long as the expected energy is finite, the theorem applies.   

Although the theorem of \cite{Eisert2002} requires one to find a non-interacting Hamiltonian with respect to which $|\psi\rangle$ has finite energy, this reference Hamiltonian need not bear any relation to the dynamics of the system of interest. Rather, the assumption of finite energy with respect to the reference merely ensures that $\rho$, which might have infinite nonzero eigenvalues, nonetheless has sufficiently accurate low-rank approximations. For instance, if we have $\mc{H}_1, \mc{H}_2 = L^2(\mathbb{R}^d)$ and one chooses $H$ to be the Hamiltonian of two independent harmonic oscillators,
\begin{align}
H = \vec{p}_1^2 + \vec{p}_2^2 + \vec{x}_1^2 + \vec{x}_2^2,
\end{align}
then any state $|\psi\rangle$ with a smooth spatial wavefunction that decays at spatial infinity at least as fast as $1/r^{(d+3)/2}$ will have finite energy with respect to $H$, and hence the theorem of \cite{Eisert2002} implies this large class of wavefunctions has finite entanglement entropy. (If one tries to weaken this condition to include wavefunctions that are not smooth but decay, or decay but are not smooth, counterexamples exist with infinite entanglement entropy in both cases.) In particular, entanglement entropy of the density matrix in Eqn.\ \ref{eqn:two-site-RDM} will be finite, as corroborated by the convergence of the numerics used for Fig. \ref{fig:entropies}. 

Similarly, the algebraic entanglement entropies of Section \ref{sec:Fock} will be finite for states with smooth, decaying wavefunctions.  The finiteness highlights the difference between the two notions of locality discussed in Section \ref{sec:axVSphix}.  Our first-quantized algebraic approach uses Type $I$ algebras and gives finite entanglement entropies, whereas the ordinary ``factorization'' of field theory gives area-law divergences, associated to the Type $III$ sub-algebras present in field theory.

\section{Discussion} \label{sec:discussion}


\subsection{Worldlines, worldsheets \& reparametrization invariance}

This work has highlighted the largely unexplored realm of target space partitions and their relevance in the quantum gravitational context. However, we remain far from our original hope of using an algebraic approach to define target space entanglement entropy in worldsheet string theory.  

In ordinary field theory, our algebraic definition successfully captured the entanglement entropy with respect to a certain factorization.  However, as discussed in Section \ref{sec:axVSphix}, the field theory admits at least two seemingly natural factorizations, which we called the ``Fock-based'' and ``field-based'' factorizations. It is the entanglement with respect to the former that is captured by our algebraic definition, whereas only the latter factorization exhibits the divergent area law contribution.  On one hand, the calculation outlined in Section \ref{sec:axVSphix} demonstrates that the additional entanglement of particle excitations atop the vacuum \textit{can} be meaningfully compared between the two factorizations.  On the other hand, the first-quantized algebraic approach remains unable to analyze the area law contribution itself.

In fact, it might appear senseless to imagine a first-quantized particle description teaching us anything about the spatial structure of the vacuum. The wordline framing of QFT suggests otherwise.
Consider the relativistic free massive scalar field.  (Interactions can be incorporated but are not the focus of the argument.) The logarithm of its partition function can be recast as the path integral of a point particle coupled to $1$-dimensional gravity on its worldline \footnote{This might be more familiar under the guise of the ``Schwinger paramterization'' of Feynman diagrams.} \cite{dijkgraaf1997houches} : 
\begin{eqnarray*}
\log Z_{QFT} & = & \log \int D\phi e^{-\frac{1}{2}\int d^{d}x\sqrt{g}\left(g^{\mu\nu}\partial_{\mu}\phi(x)\partial_{\mu}\phi(x)+m^{2}\phi(x)^{2}\right)}\\
 & = & -\frac{1}{2} \Tr \left[ \log \left(-g^{\mu\nu}\nabla_{\mu}\nabla_{\nu}+m^{2}\right)\right]\\
 & = & \int d^{d}x\bra{x^{\mu}}\int_{\epsilon}^{\infty}\frac{ds}{2s}e^{-s(p_{\mu}p^{\mu}+m^{2})}\ket{x^{\mu}}\\
 & = & \int d^{d}x\int_{y^{\mu}(0)=x^{\mu}}^{y^{\mu}(s)=x^{\mu}}\frac{Dy^{\mu}(\tau)De(\tau)}{Vol(Diff)}e^{-\frac{1}{2}\int_{0}^{s} d\tau e\left(\frac{1}{e^{2}}\partial_{\tau}y^{\mu}(\tau)\partial_{\tau}y^{\nu}(\tau)g_{\mu\nu}(y(\tau))+m^{2}\right)}
\end{eqnarray*}

This worldline approach to field theory is the most immediate field-theoretic analog of worldsheet string theory. In the worldline setting, we know we can access the area law entanglement pattern of the QFT via a replica trick Euclidean path integral. Schematically, we can compute it as 
\begin{equation}
    S_{EE}=\left( 1- n \partial_{n} ) \log Z_{QFT}[n] \right) {\bigg |}_{n=1}
\end{equation}

The right hand side of this equation, including the the necessary field-theoretic $UV$-regulator, may be completely recast in terms of worldline quantities. The euclidean path integral immediately gives us the entropy. Its Lorentzian interpretation on the other hand, remains elusive. 

Indeed, while the area law manifests itself as above in the worldline formalism, it is unclear there exists any partition of the point particle Hilbert space that yields this entropy. An algebraic approach would require such a partition.  However, hope remains. Two salient features deserve further notice. Firstly, we see the spatial arguments of the fields, the $x^\mu$ in $\phi(x^\mu )$ appear as boundary conditions on the worldline trajectories. This is the familiar statement that, in string theory, D-branes help us probe target space locality \cite{bachas1997lectures,douglas1997d}. Note the states $\ket{x^\mu }$ do not belong to the physical subspace of the point particle Hilbert space, as they do not satisfy the constraint $\hat{p}^2+m^2 \ket{\psi}=0$. In the language of \cite{dijkgraaf1997houches}, they do not reside in the BRST coholomogy of $Q_{BRST}=c(p^2 + m^2)$. This simply reflects the fact that reparametrization invariance breaks down at the endpoints of the worldline. 
Secondly, from a more algebraic perspective, we know that any reduced density matrix reproducing all two-point correlation functions  $\braket{\phi(x^\mu)\phi(y^\nu)}$, $\braket{\phi(x^\mu)\Pi(y^\nu)}$ and $\braket{\Pi(x^\mu)\Pi(y^\nu)}$ for $x^\mu , y^\nu \in A$ will gives us the field theory entanglement entropy relative to the $\phi(x)$ tensor product factorization. These correlators can also be rewritten purely in terms of  worldline variables: 

\begin{equation}
\braket{\phi(x^{\mu})\phi(y^{\nu})} =\int_{\epsilon}^{\infty}ds \bra{x^{\mu}}e^{-s(p^{2}+m^{2})}\ket{y^{\nu}}
\end{equation}

We would therefore need to consider some sort of restriction on the set of allowable ``D-branes'' for the worldline. While we have not succeeded in defining an associated reduced density matrix, it is at least clearly a Lorentzian setup.

The single particle Hamiltonian $\sqrt{\vec{p}^2 +m^2}$ considered in Section \ref{sec:axVSphix} arises via gauge fixing the relativistic point particle action. More precisely, it is the canonical Hamiltonian after choosing static gauge $x^{0}(\tau) = \tau $. We feel there is something important about reparametization invariance we have yet to pinpoint, and hope to explore this avenue in future work.

\subsection{Different factorizations and the \textit{c}=1 matrix quantum mechanics}

The discussion in Section \ref{sec:axVSphix}, teasing out the subtle differences in our notion of ``spatial locality,''  could appear somewhat artificial. Yet such competing notions of locality might, in fact, be rather generic within the emergent spacetime paradigm. The $c=1$ matrix quantum mechanics provides a sharp holographic example. 
As highlighted in \cite{seiberg2006emergent}, there exists at least two seemingly natural emergent spatial dimensions. On one hand, the matrix quantum mechanics in the singlet sector can be recast as a local fermionic field theory on matrix eigenvalue space. On the other hand, the low-energy target space dynamics, derived from the worldsheet Liouville string theory, is most naturally formulated in terms of the string embedding coordinates $X^{0}$ (the $c=1$ boson) and $\phi$ (the Liouville field).\footnote{ \cite{moore1992loops} points out important subtleties in viewing the Liouville direction as spatial coordinate. They consider instead yet another space on which they define a string field theory of loop operators, parametrized by the length of the strings they create.} 
Section 11 of \cite{sen2005tachyon} shows precisely how bosonization of the matrix model's fermionic field theory maps onto the closed string tachyon dynamics in the target spacetime. In momentum space, a simple multiplicative phase factor relates the two -  the celebrated ``leg-pole factor'' - in close parallel to the $\phi(p) \sim (2E_{p})^{-1/2} a^{\dagger}_{p}$ example discussed in Section \ref{computation}. In position space, this gives a non-local map. Natsuume and Polchinski argued all the (admittedly very simple) gravitational dynamics on the 2d target space were encoded in the matrix model via this non-local map \cite{natsuume1994gravitational}. Reference \cite{hartnoll2015entanglement} reproduced the entanglement entropy of the bulk 2d tachyon by partitioning the matrix eigenvalue space. It failed, however, in capturing any $\mc{O}(1/g_{st}^2)$  contribution - the closest 2d relative of Area/4$G$. One might blame this on choosing a notion of locality similar to the Fock space factorization discussed above, thereby capturing only excitations around the background. Making this precise might help guide future attempts at diagnosing emergent locality from matrix degrees of freedom.

\section{Acknowledgements}
It is a pleasure to thank Sean Hartnoll for stressing over the years the importance of defining spatial entanglement entropy in first-quantized (matrix) quantum mechanics. We thank him, Tom Hartman and Jordan Cotler for many fruitful discussions and early collaborations on this topic. We also gratefully acknowledge helpful conversations with Eva Silverstein and Ronak Soni. Finally, we would like to thank the Yukawa Institute for Theoretical Physics in Kyoto, where part of this work was completed during the workshop ``Quantum Information and String Theory.''

\pagebreak

\section{Appendix: The algebra for bosons} \label{sec:boson-appendix}

Here we provide more detail justifying Eqn.\  \ref{eqn:multipaticle-algebra} using the definition of the algebra $\mathcal{A}$ in Eqn.\  \ref{eqn:multiparticle-algebra-decomposition}.

First off, we can write down the commutant $\mc{A}'$ as 

\begin{equation}
    \mc{A}= \bigg \langle \{ P_{S_N} \left( \ket{\vec{x}}_{1}\bra{\vec{x}'}_1\otimes \mathds{1}_2 \otimes ... \otimes \mathds{1}_N  \right) P_{S_N} : \vec{x},\vec{x}' \in \bar{A} \} \cup \mathds{1}_{\mc{H}_{N}} \bigg \rangle  
\end{equation}
so that the center $Z(\mc{A})$ is generated by
\begin{equation}
    \mc{Z}= \bigg \langle \{ P_{S_N} \left( \int_{A} d\vec x \ket{\vec{x}}_{1}\bra{\vec{x}}_1\otimes \mathds{1}_2 \otimes ... \otimes \mathds{1}_N  \right) P_{S_N} \} \cup \mathds{1}_{\mc{H}_{N}} \bigg \rangle  
\end{equation}

At this point we wish to identify the minimal projectors which span the center $\mc{Z}$, as in Eqn.\  \ref{eqn:minimal-projectors}.  Here there are $N+1$ such projectors, which we can write as 
\begin{equation}
    \Pi_k =  \binom{N}{k} P_{S_N} \left( \underbrace{\Pi_{A} \otimes ... \otimes \Pi_{A}}_{k \: \text{times}}\otimes \underbrace{\Pi_{\bar{A}} \otimes... \otimes \Pi_{\bar{A}}}_{(N-k) \: \text{times}} \right) P_{S_N}
\end{equation}

Physically, $\Pi_k$ is the projector onto the subspace with $k$ particles in $A$ and $N-k$ particles in $\bar{A}$.

The algebra $\Pi_k \mc{A} \Pi_k $ projected onto this subspace takes the form 

\begin{equation}
    \Pi_k \mc{A} \Pi_k = \bigg \langle P_{S_{N}} \left( \ket{\vec{x}}\bra{\vec{x}'} \otimes \Pi_A...\otimes \Pi_A \otimes \Pi_{\bar{A}} \otimes... \otimes \Pi_{\bar{A}}  \right) P_{S_{N}} : \vec{x},\vec{x}'\in A \bigg \rangle 
\end{equation}

has trivial center on $\Pi_k \mc{H}_{N}$. To see this, we first write its commutant restricted to the subspace 
\begin{equation}
    \mc{A}'\big|_{\Pi_k\mc{H}_{N}}= \bigg \langle P_{S_{N}} \left( \ket{\vec{x}}\bra{\vec{x}'} \otimes \Pi_{\bar{A}}...\otimes \Pi_{\bar{A}} \otimes \Pi_A \otimes... \otimes \Pi_A \right) P_{S_{N}} : \vec{x},\vec{x}'\in \bar{A} \bigg \rangle 
\end{equation}

so that indeed the center on this subspace is trivial 

\begin{equation}
    \mc{A} \cap \mc{A}' \big|_{\Pi_k\mc{H}_{N}} = P_{S_{N}} \left ( \underbrace{\Pi_A \otimes... \otimes \Pi_A }_{k \: \text{times}} \underbrace{\Pi_{\bar{A}} \otimes... \otimes \Pi_{\bar{A}} }_{N-k \: \text{times}}  \right ) P_{S_{N}} = \mathds{1} \big |_{\Pi_k\mc{H}_{N}}
\end{equation}

Since the algebra restricted to each subspace is a factor, we know there exists a tensor product factorization in each block such that all $\mc{O}\in \mc{A}$ take the form 

\begin{equation}
    \mc{O}=\oplus_{k=0}^N \mc{O}_{A_k} \otimes \mathds{1}_{\bar{A}_{k}}
\end{equation}

What is this tensor product factorization? It is nothing but the decomposition 

\begin{equation} \label{hilbspacesum npart}
   \frac{(\mc{H}_A \oplus \mc{H}_{\bar{A}})^{\otimes N}}{S_N} = \oplus^{N}_{k=0} \left( \frac{\mc{H}_A ^{\otimes k}}{S_{k}} \otimes \frac{\mc{H}_{\bar{A}} ^{\otimes N-k}}{S_{N-k}} \right)
\end{equation}

where we define $\mc{H}_{A}^0=\mc{H}_{\bar{A}}^0=\mathbb{C}$. In particular 

\begin{equation}
    \Pi_k \mc{H}_{N} = \frac{\mc{H}_A ^{\otimes k}}{S_{k}} \otimes \frac{\mc{H}_{\bar{A}} ^{\otimes N-k}}{S_{N-k}} 
\end{equation}

Finally, we stress the symmetric projectors $P_{S_N}$ are crucial for $\mc{A}$ to contain multi-particle operators. For example, in the case of $N=2$, multiplying two basis algebra elements can generate all symmetric 2-particle operators:
\begin{eqnarray}
 P_{S_2} \left( \ket{\vec{x}}_{1}\bra{\vec{x}'}\otimes \mathds{1}_2 \right) P_{S_2}   P_{S_2} \left( \ket{\vec{y}}_{1}\bra{\vec{y}'} \otimes \mathds{1}_2 \right) P_{S_2} & = & \frac{1}{2!} P_{S_2} \ket{\vec{x}}_{1}\bra{\vec{x}'}\otimes \ket{\vec{y}}_{2}\bra{\vec{y}'} P_{S_2} \\
 & & + \frac{1}{2!} \delta(x'-y) P_{S_2} \left( \ket{\vec{x}}_{1}\bra{\vec{y}'}\otimes \mathds{1}_2 \right) P_{S_2}
\end{eqnarray}

\bibliographystyle{unsrt}
\bibliography{references}

\end{document}